\documentstyle[dina4,12pt]{article}
\newcommand{\Ms}{{\cal M}_s}
\newcommand{\erg}{{\rm erg}}
\newcommand{\cm}{{\rm cm}}
\newcommand{\bee}{\begin{equation}}
\newcommand{\ee}{\end{equation}}
\newcommand{\beea}{\begin{eqnarray}}
\newcommand{\eea}{\end{eqnarray}}
\newcommand{\R}{{\bf R}}
\newcommand{\e}{{\bf e}}
\newcommand{\rme}{{\rm e}}
\newcommand{\m}{{\bf m}}
\newcommand{\n}{{\bf n}}
\newcommand{\vx}{\vec{x}}

\newcommand{\vpi}{\vec{\pi}}
\newcommand{\tpi}{\tilde{\pi}}
\newcommand{\hV}{\hat{V}}
\newcommand{\tV}{\widetilde{V}}
\makeatletter
\@addtoreset{equation}{section}
\makeatother

\begin{document}
\begin{titlepage}
\thispagestyle{empty}
\parskip=12pt
\raggedbottom

\def\mytoday#1{{ } \ifcase\month \or
 January\or February\or March\or April\or May\or June\or
 July\or August\or September\or October\or November\or December\fi
 \space \number\year}
\noindent
\hspace*{11cm} BUTP--92/46\\
\vspace*{1cm}
\begin{center}
{\LARGE Finite size and temperature effects \\
\vspace*{0.5cm}
in the AF Heisenberg
model}\footnote{Work supported in part by Schweizerischer Nationalfonds}

\vspace{2cm}

P. Hasenfratz and F. Niedermayer\footnote{on leave from the Institute
for Theoretical Physics, E\"otv\"os University, Budapest}\\

\vspace{1cm}

Institute for Theoretical Physics \\
University of Bern \\
Sidlerstrasse 5, CH-3012 Bern, Switzerland

\vspace{1cm}

\mytoday \\ \vspace*{1cm}

\nopagebreak[4]

\begin{abstract} \normalsize

The low temperature and large volume effects in the $d=2+1$
antiferromagnetic quantum Heisenberg model are dominated
by magnon excitations.
The leading and next--to--leading corrections
are fully controlled by three physical constants, the spin stiffness,
the spin wave velocity and the staggered magnetization. Among others,
the free energy, the ground state energy, the low lying excitations,
staggered magnetization, staggered and uniform susceptibilities
are studied here. The special limits of very low temperature and
infinite volume are considered also.

\end{abstract}
\end{center}
\end{titlepage}

\section{\bf Introduction}

The low energy properties and the leading finite size and temperature
effects of the $d=2+1$ quantum anti--ferromagnetic (AF) Heisenberg model

\bee
H =  J \sum_{n,\mu} \hat{{\bf S}}_n \hat{{\bf S}}_{n+\mu}\,,
{}~~~\hat{{\bf S}}_n^2=S(S+1)
\label{1}
\ee
are governed by magnon excitations.
The interaction of massless
excitations whose existence is due to spontaneous symmetry breaking \cite{0},
is strongly constrained by symmetry principles \cite{1}.
In particle physics it was the interaction of soft pions where
many of the general features of this dynamics were first observed \cite{1,2,3}.
These investigations received new impetus recently \cite{4} by the application
of the powerful method of effective Lagrangeans \cite{2,5} which provides
a systematical way to calculate higher order corrections
(chiral perturbation theory).
Gasser and Leutwyler and their followers treated not only low energy
Goldstone boson physics but also finite temperature and finite size
effects in QCD \cite{4,6,9}. This technique can be immediately generalized
to other quantum field theories and critical
statistical systems \cite{10} and provides a systematic alternative to other
related methods \cite{14,15,16}.

The special, constrained feature of magnon interactions in the AF
Heisenberg model has been emphasized and different leading order
results were obtained in several recent publications \cite{11,12,13}.
Using current algebra techniques or intuitive reasoning it would be
very difficult to go beyond the leading order. On the other hand, this
problem becomes a relatively simple book--keeping if the technique
of chiral perturbation theory is used.
The power of chiral perturbation theory
was illustrated in this context in
ref.~\cite{17} by calculating the first correction to the correlation
length at low temperatures.
In this paper we present further results beyond leading order.
The temperature  and size dependence of the free energy density
(internal energy density, specific heat), the size dependence
of the ground state energy density, the energy of the low lying
excitations and its size dependence will be investigated.
We calculate the staggered magnetic field dependent
part of the free energy density (magnetization and susceptibility)
and study the behaviour of the uniform susceptibility.
Additionally, we discuss the zero temperature and the infinite volume limits
which reveal special physical properties and pose new technical problems.

The results of chiral perturbation theory reflect the symmetries of the
underlying model only and  depend on free physical
parameters whose number is growing  with the order of the calculation.
As it is well known, the leading order results depend
on three physical parameters: the spin stiffness constant $\rho_s$,
the spin wave velocity $c$
and the staggered magnetization $\Ms$.
It is less well known, perhaps,
that for the quantities we study here the next--to--leading
corrections can be expressed in terms of the same constants,
no new parameters enter (see ref.~\cite{10} and Section~4).
As we shall discuss, these predictions are free of cut--off effects also.

The three parameters $\rho_s$, $c$ and $\Ms$ are fixed by the underlying
model as the function of the coupling $J$. An effective way to fix these
constants is to compare the results of chiral perturbation theory with
the corresponding numerical results using the Hamiltonian operator
in eq.~(\ref{1}). This method has been applied earlier for classical
ferromagnetic models and field theories  \cite{18}.

The classical O($3$) ferromagnetic Heisenberg model in $d=3$ in the
broken phase is described also by the results  of this paper.
In this case we have to
put $c=1$ and interpret $L_t=T^{-1}$ and $L_s$ as giving the size
of the euclidean box $L_t \times L_s \times L_s$ on which the
model is defined.
Actually, we give all the results for the symmetry group O($N$),
$N=3$ corresponding to the model in eq.~(\ref{1}).

The paper is organized as follows. For the readers convenience we
summarize the results in Section~2. Concerning the notations, results
and region of validity, this section is self--contained.
Some of the results discussed in Section~2 occurred earlier in the
literature in the context of classical ferromagnets or field theories.
In this case our role is only to insert the proper spin wave velocity
dependence and to raise attention to the existence of these results.
In Section~3 we try to clarify some problematic results
which occur in the literature. In Section~4 we construct the effective
Lagrangean which is needed to derive the free energy density
$f=f(T,L_s)$ up to next--to--leading order. The effect of a finite
cut--off is discussed in some detail there also.
In Section~5 we calculate $f(T,L_s)$ up to next--to--leading order.
Although this calculation is a priori valid for ${\hbar c  / (TL_s)}=$O($1$)
only, the $L_s \to \infty$ limit gives correctly the temperature dependence
of the infinite volume free energy density at low temperatures.
The formal $T \to 0$ limit, however,
does not lead to the correct volume dependence of the ground state
energy, as we discuss in Section~5.
The volume dependence of the ground state energy, the lowest lying
excitations and their finite volume corrections are considered
in Section~6. In this section we regularize the effective Lagrangean
on a lattice which opens the possibility to discuss the issue of power
divergences in chiral perturbation theory. We show that
--- as expected --- all these divergences can be absorbed through the
standard renormalization procedure and no mixing between infrared and
ultraviolet problems occurs.
In Section~7 the dependence of the free energy density on a weak staggered
magnetic field, the magnetization and susceptibility are considered.
Since this problem has been treated before we can keep this part brief.
Very low temperatures require special care again as discussed in Section~8,
where the volume dependence of the staggered susceptibility at $T=0$
is calculated. In Section~9 the uniform susceptibility is discussed.
In the Appendix A we consider specific lattice momentum sums and integrals.
Appendix B contains some asymptotic expressions for the shape coefficients.

Chiral perturbation theory is an exact low energy expansion.
It summarizes the symmetry properties of the model in eq.~(\ref{1})
in a concised way and it is valid for any spin $S$. Consequently, if we
replace in any quantity considered here the parameters $\rho_s$, $c$
and $\Ms$ by their spin wave  ($1/S$) expansion, then the spin wave
expansion of this quantity is obtained. The relation between these
two expansions is discussed further in Section~3.

We made an effort to stay close to the notations used earlier in the
literature. Since the model in eq.~(\ref{1}) gave the motivation
for this work,
we denoted by $\rho_s$ (rather than by $F^2$, or $\Upsilon$)
the leading non-magnon scale.
We used the index $s$ both to denote staggered quantities and
to refer to 'spatial', like spatial volume and size. Certain parts
of the calculation are carried out for general dimension $d$, in which
case we kept $d$ as a parameter in the formulas. In the calculations
we used the convention "$\hbar=c=1$" and restored the spin wave velocity
dependence only at the end. At some places we introduced more
convenient notations with respect to the preprint version of
this paper (BUTP--92/46).

\section{\bf  Results}

Let us introduce a staggered external field ${\bf B}_s=(B_s,0,0)$,
and a uniform magnetic field\footnote{The magnetic moment is included
in the definition of $h_u$}
${\bf h}_u=(h_u,0,0)$ and generalize eq.~(\ref{1}) to
\bee
H =  J \sum_{n,\mu} \hat{{\bf S}}_n \hat{{\bf S}}_{n+\mu}
-{\bf B}_s \sum_n (-1)^{\| n \|} \hat{{\bf S}}_n
-{\bf h}_u \sum_n  \hat{{\bf S}}_n \,.
\label{2}
\ee
We shall define the field ${\bf H}_s$ as

\bee
{\bf H}_s =  {{\bf B}_s \over a^2}\,,
\label{3}
\ee
where $a$ is the lattice unit of the square lattice over which
eq.~(\ref{2}) is defined. We shall consider a system of spatial
size $L_s \times L_s$ at temperature $T$. The spin--stiffness,
the spin wave velocity and the staggered magnetization of the
infinitely large system  at zero temperature and at zero external fields
are denoted by  $\rho_s$, $c$ and $\Ms$, respectively.
These quantities are fixed by the Hamilton operator in eq.~(\ref{2}).
On dimensional grounds $\rho_s$ and $\hbar c/a$ are proportional to $J$.
It will be useful to introduce the notation
\bee
l^3={\hbar c \over TL_s} \,.
\label{4}
\ee
In the path integral formulation we shall work in a euclidean box
$L_t \times L_s \times L_s$, where $L_t=\hbar c/T$. We shall refer
to the $l=O(1)$ and $l\gg 1$ cases as "cubic" and "cylinder" geometry,
respectively.
Concerning dimensions, we have
\beea
& & {\rm dim}\; J=\erg,~~{\rm dim}\; H_s={\erg \over \cm^2},
    ~~{\rm dim}\; L_s=\cm,~~ {\rm dim}\; T=\erg, \nonumber \\
& & {\rm dim}\; h_u=\erg,~~ {\rm dim}\; \rho_s=\erg,~~{\rm dim}\;
 \Ms=1,~~{\rm dim}\; l=1 \,.
\label{5}
\eea
The free energy density $f$ is defined as
\bee
Z = \exp\left( -{L^2_s \over T} f \right)\,,
\label{6}
\ee
where Z is the partition function. As explained in the introduction,
the following results apply to classical O($N$) symmetric ferromagnetic
systems also, while eq.~(\ref{2}) corresponds to $N=3$.
In general, magnons dominate the physical properties of the system
at low energies, large volumes, low temperatures and weak external
fields. More detailed conditions will be given together with the results.

\subsection{{\bf  $H_s$--dependence of the basic parameters}
($L_s=\infty$, $T=0$, $H_s \neq 0$, $h_u=0$)}

In the presence of a nonzero staggered field $H_s$ the magnons
(they have a relativistic dispersion relation) pick up a mass $M^H$.
Let us introduce the notation $M$:
\bee
(c^2 M)^2 = {(\hbar c)^2 \Ms \over \rho_s} H_s \,.
\label{7}
\ee
We shall consider weak external staggered fields such that
\bee
c^2 M  \ll  \rho_s \,.
\label{8}
\ee
Denoting the staggered magnetization and the spin stiffness in the presence
of the field $H_s \neq 0$ by $\Ms^H$ and $\rho_s^H$, we have
\beea
\Ms^H & = & \Ms \left[ 1+{N-1 \over 8\pi}\cdot{c^2 M \over \rho_s}+
         {\rm O}(H_s) \right] \,, \nonumber \\
\rho_s^H & = & \rho_s \left[ 1+2\cdot{N-2 \over 8\pi}\cdot{c^2 M \over \rho_s}+
          {\rm O}(H_s) \right] \,, \label{9} \\
M^H & = & M \left[ 1- {1\over 2}\cdot {N-3\over 8\pi}\cdot{c^2 M \over \rho_s}+
         {\rm O}(H_s) \right] \,. \nonumber
\eea
Actually, these are first order results and --- at least in the context
of ferromagnetic systems --- are well known  \cite{14,4,10}.
We study these relations in Section~6 in order to fix the renormalization
of the parameters.

\subsection{{\bf  The free energy in "cubic" geometry}
($L_s={\rm finite}$, $T \neq 0$, $H_s =0$, $h_u=0$, $l={\rm O}(1)$)}

We consider large volumes and low temperatures relative to typical
non--magnon scales:
\bee
{\hbar c \over L_s} \ll \rho_s \,,~~ T \ll \rho_s \,,
\label{10}
\ee
while keeping the ratio $l$ finite, O($1$). The free energy density reads
\beea
f(T,L_s) & =  & q_1 +{T \over L_s^2} q_2
- {N-1 \over 2} {T \over L_s^2}\left\{  {1\over 3}\ln{L_s^2 \over T}
+\beta_0(l) \right.  \label{11} \\
 &  & +\left. (N-2) {\hbar c \over \rho_s L_s l}\beta_1(l)
+{\rm O}\left( {1 \over  L_s^2} \right) \right\}
\,.  \nonumber
\eea
Here $q_1$ and $q_2$ are two non--universal constants which are not fixed
by chiral perturbation theory. The constant $q_1$ is the infinite volume
ground state energy density which is studied intensively for the model
in eq.~(\ref{1}) numerically \cite{28a,21,19,29}
and otherwise \cite{28b,23,24,25}.
(For a detailed discussion, see refs.~\cite{28,Man}.)
The constant $q_2$ depends
on the precise normalization of the partition function. The internal
energy density is already free of $q_2$.
The shape coefficients $\beta_n(l)$  are expressed through
$\alpha_n(l)$ by the relations \cite{10}
\beea
& & \beta_0(l)= \alpha_0(l)+\ln (4\pi)-\gamma_E-{2\over 3}=
\alpha_0(l)+1.287142\ldots \,, \nonumber \\
& & \beta_n(l)=\left(-{1 \over 4\pi} \right)^n \left[ \alpha_n(l)+
{3 \over n(2n-3)} \right] \,,~~~n=1,2,\ldots \,. \label{14a}
\eea
The functions $\alpha_p(l)$ can be calculated easily using the following
representation  \cite{10}
\bee
\alpha_p(l)={\hat \alpha}_{p-3/2}(l)+{\hat \alpha}_{-p}\left(
{1\over l} \right) \,,
\label{12}
\ee
where
\bee
{\hat \alpha}_r(l)= \int_0^1 dt t^{r-1}\left[ S \left({1\over l^2t}\right)^2
S\left({l^4\over t} \right) -1 \right] \,,
\label{13}
\ee
with
\bee
S(x)=\sum_{-\infty}^{\infty} \rme^{-\pi n^2 x}\,.
\label{14}
\ee
For $l=1$ the relevant $\beta_n$ values are given in Table 1 of
ref.~\cite{10}. In Appendix B we give approximate analytic forms,
which are sufficiently precise for applications.

The internal energy density
$e(T,L_s)=-T^2 {\partial \over \partial T} \left(f(T,L_s) /T \right)$
can be written in the form
\beea
e(T,L_s) & = & q_1-{N-1 \over 6}{T\over L_s^2}
\left\{ 1+l{d\over dl}\beta_0(l)\right. \nonumber \\
&  & - \left. (N-2){\hbar c\over \rho_s L_s l}
\left(\beta_1(l)-l{d\over dl}\beta_1(l)\right)+\ldots \right\}
\,.
\label{15}
\eea
In ref.~\cite{17} we quoted this result for $N=3$ with a slight
error\footnote{Neither the text
nor the subsequent equations are influenced by this error in \cite{17}.},
unfortunately: in the $T^3$ coefficient in the brackets the
number "1" was incorrectly given as "2".

\subsection{{\bf Thermodynamics at infinite volume}
($L_s=\infty$, $T \neq 0$, $H_s =0$, $h_u=0$)}

Although the results in 2.2 were derived under the condition
$l=$O($1$), the $L_s \to \infty$ ($l \to 0$) limit is
smooth\footnote{see Appendix B}
and gives correctly the temperature dependence of the free energy and
internal energy densities:
\beea
f(T) & = & q_1-(N-1) {\zeta (3) \over 2\pi (\hbar c)^2} T^3
   +{\rm O}(T^5) \,, \nonumber \\
e(T) & = &  q_1+(N-1) {\zeta (3) \over \pi (\hbar c)^2} T^3
   +{\rm O}(T^5)
\,.  \label{16}
\eea
Here $\zeta(3)=1.2020569$. There is no $\sim T^4$ contribution.
Eq.~(\ref{16}) gives for the specific heat
\bee
C(T)=(N-1){3\zeta(3) \over \pi(\hbar c)^2} T^2 +{\rm O}(T^4)\,.
\label{17}
\ee
The $\sim T^3$ term in eq.~(\ref{16}) is just the free energy
contribution of $N-1$ massless free bosons. This prediction
has occurred earlier in  \cite{13} for the AF Heisenberg model ($N=3$).
Although our calculation goes up to order $\sim T^4$, the corresponding
coefficient turns out to be zero.
There is an amazing mess in the literature on the $\sim T^3$ term
in the internal energy density of the AF Heisenberg model.
We shall return to this problem in Section~3.

\subsection{{\bf Volume dependence of the ground state energy density}
($L_s={\rm finite}$, $T=0$, $H_s=0$, $h_u=0$)}

Eq.~(\ref{11}) was derived under the condition $l=$O($1$) and
as we shall discuss later, the $T \to 0$ ($l \to \infty$) formal
limit of eq.~(\ref{11}) does {\em not} give the correct volume
dependence of the ground state energy density.
Very low temperatures require special care.
We give two different derivations leading to
\bee
\epsilon_0(L_s)=q_1-{N-1 \over 2}\cdot{1.437745(\hbar c) \over L_s^3}
+{(N-1)(N-2)\over 8}\cdot{(\hbar c)^2 \over \rho_s L_s^4}+
{\rm O}\left({1 \over  L_s^5} \right) \,.
\label{18}
\ee
The $\sim 1/L_s^3$ correction has been derived earlier for the
AF Heisenberg model ($N=3$) using current algebra like techniques \cite{12}
and effective Lagrangeans \cite{13}.
In some works  \cite{21} the $\sim 1/L_s^4$ contribution to the ground state
energy density was predicted to be zero, which is incorrect.
The problem of the ground state energy has been discussed in the context
of QCD in $d=3+1$ with SU($n$)$\times$SU($n$) symmetry  \cite{7}.
Since SU($2$)$\times$SU($2$)$\sim$O($4$), eq.~(\ref{18}) for $N=4$ should
agree with the corresponding result for $n=2$ in  \cite{7}.
After trivial adjustments corresponding to $d=2+1~\to~d=3+1$ they agree,
indeed.

\subsection{{\bf The energy of the lowest lying excitations and its
volume dependence} ($L_s={\rm finite}$, $T=0$, $H_s=0$, $h_u=0$)}

It is known since a long time that the tower of lowest lying excitations
in the broken phase is related to the slow precession of the order
parameter and, in leading order, it is given by simple quantum mechanics
over the symmetry group space  \cite{15,22,7}.
We have calculated the leading finite size correction also and
obtained for the energy difference between the excited state with
$j=1,2,\ldots$ and the ground state\footnote{Eq.~(\ref{19}) refers to the
energy and not to the energy density: $E_0(L_s)=L_s^2 \epsilon_0(L_s)$.}
\bee
E_j(L_s)-E_0(L_s)=j(j+N-2){(\hbar c)^2 \over 2 \rho_s L_s^2}
\left[ 1-(N-2){\hbar c \over \rho_s L_s} {3.900265\over 4\pi}
+{\rm O}\left( 1 \over  L_s^2 \right) \right] \,.
\label{19}
\ee
For the AF Heisenberg model ($N=3$) the leading order result has been
discussed in  \cite{12}.

\subsection{{\bf Weak staggered field in "cubic" geometry}
($L_s={\rm finite}$, $T \neq 0$, $H_s \neq 0$, $h_u=0$, $l={\rm O}(1)$)}

We quote now the $H_s$--dependent part of the free energy density derived
under conditions of eq.~(\ref{10}) and\footnote{With the $\le$ sign in
eq.~(\ref{20}) we want to indicate that $H_s$ can be arbitrarily small.}
\bee
u_0 \equiv {H_s \Ms L_s^2 \over T} \le {\rm O}(1) \,.
\label{20}
\ee
This result was derived earlier  \cite{10}, we restored the correct spin wave
velocity dependence only:
\bee
f(T,L_s,H_s)=f(T,L_s)-
{T\over L_s^2}\left[ \ln \left(Y_N(u)\right) +\gamma_2 \cdot
\left({u_0\hbar c \over \rho_s L_s l}\right)^2 +\ldots \right] \,,
\label{21}
\ee
where the $H_s$ independent part, $f(T,L_s,H_s=0)\equiv f(T,L_s)$
is given by eq.~(\ref{11}).
We also introduced the notations\footnote{We have changed the notation
$\rho_1$, $\rho_2$ of \cite{10} to $\gamma_1$, $\gamma_2$ to
avoid confusion with the spin stiffness.}
\beea
& & u=\gamma_1 u_0   \,, \nonumber  \\
& & \gamma_1 = 1+{N-1\over 2}{\hbar c\over \rho_s L_s l} \beta_1(l)
-{(N-1)(N-3)\over 8}\left( {\hbar c\over \rho_s L_s l}\right)^2
\left(\beta_1(l)^2-2\beta_2(l)\right)
\,, \nonumber  \\
& & \gamma_2={N-1 \over 4} \beta_2(l)  \,,  \label{22}
\eea
and $\beta_p(l)$ was defined earlier in eqs.~(\ref{14a}--\ref{14}).
The function $Y_N(z)$ can be expressed in terms of the modified
Bessel function $I_{\nu}(z)$,
\bee
Y_N(z)= \left( {z \over 2} \right)^{-{N-2 \over 2}}I_{{N-2 \over 2}}(z)\,.
\label{23}
\ee
Some of the properties of the function $Y_N(z)$ are summarized in
Appendix F of ref.~\cite{10}.

Derivatives with respect to the staggered field $H_s$ give the staggered
magnetization and susceptibilities,
\bee
\Ms(T,L_s,H_s)=-{\partial \over \partial H_s} f(T,L_s,H_s)=
\Ms \left\{ \gamma_1 {{Y'}_N (u) \over Y_N(u)}
+2 \gamma_2 u_0 \left( \hbar c \over \rho_s L_s l \right)^2 +\ldots\right\} \,,
\label{24}
\ee
\beea
& & \chi_s^{\|}(T,L_s,H_s)   \equiv
{\partial \Ms(T,L_s,H_s)\over\partial H_s} =
a^2 {\partial \Ms(T,L_s,H_s)\over\partial B_s} \nonumber \\
 & & ~~~~~~ =  \Ms^2 {L_s^2 \over T} \left\{ \gamma_1^2 \left[
{{Y''}_N(u) \over Y_N(u)}-\left( {{Y'}_N(u) \over Y_N(u)}\right)^2\right]
+2\gamma_2 \left( \hbar c \over \rho_s L_s l \right)^2 +\ldots\right\} \,,
\label{25} \\
 & & \chi_s^{\bot} (T,L_s,H_s)  =  {\Ms(T,L_s,H_s) \over H_s} \,.
\nonumber
\eea
In the $H_s \to 0$ limit $\Ms(T,L_s,H_s) \to 0$ as it should, and
\beea
 \chi_s^{\|} (T,L_s,0)  & = & \chi_s^{\bot} (T,L_s,0)  =
{1 \over N} \Ms^2 {L_s^2 \over T} \left\{
1+(N-1)\left( {\hbar c \over \rho_s L_s l}\right)\cdot \beta_1(l)\right.
\nonumber \\
& & \left. +{N-1 \over 2} \left({\hbar c \over \rho_s L_s l} \right)^2
\left( \beta_1(l)^2 +(2N-3)\beta_2(l)\right)+\ldots \right\} \,.
\label{26}
\eea

\subsection{Volume dependence of the staggered
susceptibility at small temperature
 ($T \ll \hbar c / L_s$, $H_s=0$, $h_u=0$)}

As we mentioned before, low temperatures require special treatment.
For $T \ll \hbar c / L_s$ the coefficients  multiplying the expansion
parameter $\hbar c /\rho_s L_s$ in eq.~(\ref{26}) become large.
The dynamics is dominated by the low lying modes of the effective O($N$)
rotator. The corresponding susceptibility is calculated in Section 8.
We note here only that for temperatures
\bee
{TL_s^2\rho_s \over (\hbar c)^2} \ll 1 \,,
\label{27a}
\ee
the susceptibility becomes independent of the temperature and it is
proportional to the square of the spatial volume:
\bee
 \chi_s (T,L_s)  = {4\over N(N-1)}{\Ms^2 \rho_s\over (\hbar c)^2}
L_s^4 \left[ 1+(2N-3){\hbar c\over \rho_s L_s}{3.900265\over 4\pi}
+{\rm O}\left({1\over L_s^2}\right) \right] \,.
\label{27b}
\ee
The basic reason of this strong volume dependence lies in the smallness
of the excitation energies which are proportional to the inverse spatial
volume, eq.~(\ref{19}).

\subsection{Uniform susceptibility ($H_s=0$, $h_u=0$)}

For cubic geometry, $T={\rm O}(\hbar c/L_s)$ and under the conditions
in eq.~(\ref{10}) we obtain for the uniform
susceptibility at $h_u=0$
\beea
\chi_u (T,L_s)
& =& {2\over N} {\rho_s \over (\hbar c)^2}
 \left\{ 1+{N-2\over 3}{\hbar c\over \rho_s L_s l}
\tilde{\beta}_1(l)  \right. \label{27d} \\
& &+\left. {N-2\over 3} \left({\hbar c\over \rho_s L_s l}\right)^2
 \left[(N-2)\tilde{\beta}_2(l)
-{1\over 3} \tilde{\beta}_1(l)^2 -6\psi(l)
\right] \right\} \,. \nonumber
\eea
Here we introduced the shorthand notations
\bee
\tilde{\beta}_1(l)= {1\over l} {d\over dl}\left(l^2\beta_1(l)\right)\,,~~~
\tilde{\beta}_2(l)= {1 \over l^3} {d\over dl}\left(l^4\beta_2(l)\right)\,.
\label{27e}
\ee
In the next--to--leading order a new shape coefficient $\psi(l)$
enters, whose definition is given in Section 9. For $l=1$
it can be expressed through $\beta_n$,
\bee
\psi(1)=-{1\over 3} \beta_1(1)^2-{1\over 3} \beta_2(1)
=-0.020529\,,
\label{27f}
\ee
for other values of $l$ it requires a less trivial numerical
evaluation.
In Appendix~B we give approximate analytic expressions for $\psi(l)$.

In the infinite volume limit eq.~(\ref{27d}) gives
\bee
\chi_u(T)={2\over N}{\rho_s\over (\hbar c)^2}
\left[ 1+(N-2){T\over 2\pi\rho_s}+(N-2)\left( {T\over 2\pi\rho_s}\right)^2
+\ldots \right] \,. \label{27g}
\ee
Here $\chi_u(T)$ is the angular average of the transversal and longitudinal
susceptibilities.  This is reflected in the result
$\chi_u(T\to 0) =(2/N)\rho_s/(\hbar c)^2 =(2/N)\chi_u^{\bot}$.

For $T \ll \hbar c /L_s$ the result is again given by the effective
O($N$) rotator. In particular for very small temperatures
$TL_s^2\rho_s /(\hbar c)^2 \ll 1$ the susceptibility becomes exponentially
small
\bee
\chi_u (T,L_s)={2\over N} {1\over L_s^2 T}
\exp\left\{-{N-1\over 2}{(\hbar c)^2\over \rho_s L_s^2T} \right\} \,.
\label{27c}
\ee

\section{\bf  Chiral perturbation theory vs. spin wave expansion.
Discussion on some problematic results in the literature}

Chiral perturbation theory is an exact low energy expansion.
The basic condition for the validity of this expansion is that the scales
where the magnon physics is probed (like $T$ or $\hbar c/L_s$) should
be much smaller than the typical non--magnon scales (like $\rho_s$).
In eqs.~(\ref{11},\ref{18}), for example, $\hbar c / \rho_s L_s$
enters as an expansion parameter. For large spin $S$,\footnote{We
consider $N=3$ and discuss the AF Heisenberg model in this section.}
$\rho_s\sim JS^2$, while $\hbar c \sim JS$, therefore the expansion parameter
is $\hbar c /\rho_s L_s \sim 1/S$. Since the spin wave expansion is an
$1/S$ expansion, for certain quantities the two expansions show analogies.
The differences are more important, however. Consider eq.~(\ref{16}),
for example. This result says that the internal energy of the AF Heisenberg
model satisfies
\bee
\lim_{T\to 0} {e(T)-e(0) \over T^3} = {2 \zeta(3) \over \pi (\hbar c)^2}\,,
\label{27}
\ee
which is an {\em exact} statement. There are no corrections to this result
from higher order chiral perturbation theory or from other sources.
Spin wave expansion gives only a $1/S$ expansion for the $\sim T^3$
coefficient. On the other hand, as it is discussed in the introduction,
eq.~(\ref{27}) is a symmetry relation only, it is satisfied for any $S$.
Consequently, replacing the left and right hand sides of eq.~(\ref{27}) by
the corresponding spin wave expansion one obtains an identity.
This might serve as a good check on the results.

As a further example consider the temperature dependence of the uniform
susceptibility in the large volume limit, eq.~(\ref{27g}). By replacing
$\rho_s$ and $\hbar c$ in it by their spin wave expansion form,
which is available up to O($1/S^2$) \cite{Iga}, one obtains the spin wave
expansion for the uniform susceptibility. This result is consistent with
a direct spin wave expansion for $\chi_u(T)$ up to the known order
\cite{24}.

In the following we would like to discuss some problematic results
in the AF Heisenberg model.

\subsection{\bf The leading temperature dependence of the internal energy}

The exact result, as we discussed before, is given by eq.~(\ref{27}).
Leading spin wave expansion predicts $\hbar c =\sqrt{2} Ja$ which
gives on the right hand side of eq.~(\ref{27}) $0.3826/(Ja)^2$.
This is in agreement with the leading spin wave result for the left hand
side  \cite{23}. Including next--to--leading corrections one obtains  \cite{24}
$\hbar c = \sqrt{2} Ja \cdot 1.158$. Inserting this number in
eq.~(\ref{27}), the $\sim T^3$ coefficient is predicted to be
$0.2853/(Ja)^2$. This is in agreement with Takahashi's
modified\footnote{Takahashi's result corresponds to inserting $\hbar c$
at $S=1/2$ into the denominator in eq.~(\ref{27}) rather than
expanding the expression in $1/S$.} spin wave theory,
eq.~(26) in  ref.~\cite{24}.

These results are consistent. Schwinger boson mean field theory  \cite{25}
leads to the same equations as the modified spin wave theory in  \cite{24}.
These equations were studied numerically in  \cite{25} and
$(0.77 \pm 0.03) /(Ja)^2$ was predicted for the right hand side of
eq.~(\ref{27}).  This is almost a factor 3 larger than the correct
result and is due, presumably, to some numerical problems in
solving the equations  \cite{26}. In  \cite{27} this incorrect
result was confirmed in a
Monte Carlo calculation. The reason might be that the simulation did not reach
sufficiently low temperatures to be compared with a $C(T) \sim T^2$
behaviour. In  \cite{28}, the modified spin wave result for $e(T)$ was
incorrectly quoted as $2\cdot 0.2853/(Ja)^2\cdot T^3$.
This factor of $2$ influenced some related conclusions in ref.~\cite{28}.

\subsection{\bf The O($1/L_s^4$) correction to the ground
state energy density}

As we discussed in Section 2.5, the lowest excitations are described
by a quantum mechanical rotator giving a $\sim j(j+1)\cdot 1/L_s^4$
contribution
to the energy density. Although this term is zero for $j=0$ (ground state),
there are non--zero $\sim 1/L_s^4$ corrections to the ground state energy
density from other fluctuations (see eq.~(\ref{18})).
This term is missing in earlier discussions and analyses  \cite{21,28,29}.
For $N=3$ we can write
\bee
\epsilon_0(L_s)=q_1-{1.4377 (\hbar c) \over L_s^3}
\left[ 1- {\hbar c \over 5.7508 \rho_s} \cdot {1 \over L_s} \right]
+{\rm O} \left( {1 \over L_s^5} \right) \,.
\label{28}
\ee
If we take $\rho_s /J \approx 0.2$, $\hbar c /J a \approx 1.6$  \cite{28,Man},
the correction is significant for $L_s /a \in (4,\ldots,12)$ where the
best numerical results are available.
A $1/L_s^3$ fit might lead to a significantly distorted spin wave velocity.

\subsection{\bf Uniform magnetic field at infinite volume and at
low temperatures}

The uniform external field is coupled to a conserved quantity in the
AF Heisenberg model. It enters the Hamilton operator like a chemical potential.
In the effective Lagrangean this chemical potential shows up like the
$t$--component of a constant imaginary non--Abelian gauge field.
Here $t$ is the third ("temperature") direction in the
$L_s \times L_s \times (\hbar / T)$ euclidean box.

Switch off now the magnetic field and take $L_s=\infty$, $T \ll \rho_s$.
As it is observed before, new non--perturbative effects occur in this
limit, which can be related to the $d=2$ classical
non--linear sigma model \cite{11,17}.
The $\infty\times\infty\times (\hbar / T)$
slab is reduced to an $\infty\times\infty$ two dimensional model.
The intuitive reason is that the correlation length along the spatial
dimensions is exponentially large in $1/T$, so the width of the slab
($\hbar/T$) becomes very small compared to the correlation length.
In the $d=2$ non--linear sigma
model it is an excellent technical tool to probe the system under a chemical
potential corresponding to a constant gauge field along one of the two
spatial directions \cite{32,17}.
Among others, exact results can be obtained for
the chemical potential dependence of the free energy.
This chemical potential has, however, no direct relation to the uniform
magnetic field in the AF Heisenberg model,
which corresponds to a constant gauge field
along the "temperature" direction. This fact modifies the conclusions of
a recent paper \cite{33} significantly.

\bigskip
\section{\bf The effective Lagrangean. Discussion on the cut--off effects}

In this section we construct the effective Lagrangean which serves to derive
the free energy density up to next--to--leading order. Since for quantum field
theories or critical statistical systems this problem has been discussed
in detail before \cite{4,6,8,10}, we can be brief on those points.

The AF quantum Heisenberg model in eq.~(\ref{1}) has two parameters:
the coupling $J$ which sets the scale and the value of the spin $S$ which
takes discrete values. The smaller is $S$, the closer is the system to the
boundary between the broken and symmetric phases. It is rigorously established
that the ground state is ordered for $S \ge 1$ \cite{34}.
Approximate analytic and
numerical results suggest that $S=1/2$ is in the broken phase also \cite{35}.
For every $S$ we have two massless excitations (magnons) with a
relativistic dispersion relation. This is similar to the case of a continuum
quantum field theory, or a classical ferromagnet in the critical region
with a broken O($3$) symmetry. On the other hand, in the later cases the
non--Goldstone boson scales (e.g. $\rho_s$) are also much smaller
than the cut--off. This is not so in the model in eq.~(\ref{1}).
Take, for example the length scale corresponding to the spin stiffness,
$\xi_{\rho}=\hbar c / \rho_s$. For an order of magnitude estimate we can
take $\hbar c =1.6\cdot Ja$, $\rho_s=0.2 \cdot J$ \cite{28,Man} giving
$\xi_{\rho}/a=8$, as opposed to a critical system where this ratio goes
to infinity. In general, this cut--off dependence modifies the effective
prescription.

In the method of effective Lagrangeans one writes down the most general
local Lagrangean $L_{eff}$ (in terms of the smooth field corresponding
to the order parameter) which respects the symmetries of the underlying
model. The natural --- although strictly unproved --- expectation is
that the predictions from $L_{eff}$ reflect the symmetries only
and correspond to any underlying model having the same kind of Goldstone
bosons (magnons) and symmetries. This method becomes practically interesting
through the observation that Goldstone bosons (magnons) interact weakly
at low energies and a systematic perturbative expansion can be advised.
In any given order only a finite number of unknown low energy constants
enter.

Forget about the cut--off effects first and consider the problem of the free
energy density $f(T,L_s)$ for small $T$ and $L_s^{-1}$ under the condition
$l=$O($1$) where $l$ is defined in eq.~(\ref{4}).
Let us take "$\hbar c =1$" in the following discussion, the missing
$\hbar c$ factors can be inserted at the end using dimensional analysis.
The effective Lagrangean is constructed as a general non--linear
$\sigma$--model in terms of the field variables ${\bf S}$, ${\bf S}^2=1$:
\beea
& & {\bf S}(x)=\left( \left( 1- \vec{\pi}^2(x) \right)^{1/2}, \vec{\pi}(x)
\right) \,, \label{29} \\
& & \vec{\pi}(x)=\left( \pi^1(x),\pi^2(x) \right),~~
x_{\mu}=(x_0 = t,x_1,x_2) \,. \nonumber
\eea
The low energy excitations carry momenta $p\sim L_s^{-1} \sim T$.
Then, in the effective Lagrangean, every derivative $\partial_{\mu}$
is counted as $\sim p$. The terms in the effective model should comply
with the symmetries (most notably with the O($3$) symmetry)
of the underlying theory in eq.~(\ref{1}). The different terms
in the effective Lagrangean are multiplied with unknown couplings.
The leading term in the effective action contains the least number of
derivatives\footnote{repeated indices are summed},
\beea
{\cal A}&=&\int_0^{1/T} dt \int_0^{L_s}d^2x \, {1\over 2} \rho_s
\partial_{\mu}{\bf S}(x)\partial_{\mu}{\bf S}(x) \label{30} \\
 &=& \int_0^{1/T} dt \int_0^{L_s}d^2x \, {1\over 2} \rho_s
\left\{ \partial_{\mu}\vec{\pi}\partial_{\mu}\vec{\pi}+
\frac{(\vec{\pi}\partial_{\mu}\vec{\pi})(\vec{\pi}\partial_{\mu}\vec{\pi})}
{1-\vec{\pi}^2} \right\}, ~~~\mu=0,1,2 \nonumber
\eea

In eq.~(\ref{30}), we denoted the coupling by $\rho_s$
using the well known result that for $T=0$, $L_s=\infty$ this
coupling is the spin stiffness (helicity modulus, square of the Goldstone
boson decay constant) and the fact that the couplings of the effective
Lagrangean are independent of $T$ and $L_s$ \cite{8}.\footnote{This is true if
periodic boundary conditions are used which we shall assume in the following.}
The field $\vec{\pi}$ should be counted as $\sim p^{1/2}$, since fluctuations
of this size have a Boltzmann factor of O($1$).
Indeed, the leading term in the Lagrange density in eq.~(\ref{30}) is
\bee
{1\over 2} \rho_s \partial_{\mu}\vec{\pi}\partial_{\mu}\vec{\pi}
\sim p^3 \,,
\label{31}
\ee
which is integrated over a region $ L_s^2 /T \sim 1/p^3$.
The leading term in eq.~(\ref{31}) gives an O($p^3$) contribution to the
free energy density. The terms obtained by expanding the denominator in the
second term in eq.~(\ref{30}) give contributions of O($p^4$), O($p^5$),
\ldots . Of course, there are other terms in the general effective Lagrangean
which can also contribute in higher order. The terms with four derivatives
have the form
\bee
\frac{1}{2} g_4^{(1)}
\left( \partial_{\mu} \partial_{\mu} {\bf S}
 \partial_{\nu} \partial_{\nu} {\bf S} \right)+
\frac{1}{4} g_4^{(2)}
\left( \partial_{\mu}  {\bf S} \partial_{\mu} {\bf S} \right)^2+
\frac{1}{4} g_4^{(3)}
\left( \partial_{\mu}  {\bf S} \partial_{\nu} {\bf S} \right)
\left( \partial_{\mu}  {\bf S} \partial_{\nu} {\bf S} \right) \,.
\label{32}
\ee
\smallskip
By a change of integration variable in the path integral the first term
in eq.~(\ref{32}) can be transformed away \cite{10}, so we can take
$g_4^{(1)}=0$. As it is easy to see, the other terms in eq.~(\ref{32})
are at least O($p^6$). Terms in the effective Lagrangean containing more
then four derivatives are even more suppressed. We can conclude
therefore that if the cut--off effects can be neglected, the effective action
in eq.~(\ref{30}) determines the O($p^3$), O($p^4$) and O($p^5$) parts
in the free energy density. Only two unknown parameters, $\rho_s$ and
--- after restoring dimensions --- $c$, enter up to this order.

Let us now discuss the cut--off effects.
The effective prescription is influenced  by a finite cut--off in two ways.
First, the finite cut--off might break some of the symmetries
(for example, space rotation symmetry) which the underlying model would have
in the continuum limit. This results in new terms in the effective Lagrangean
which at infinite cut--off would be excluded. Second, the Fourier
integrals (sums) of the chiral perturbation theory will depend on
a finite cut--off.
The divergent cut--off dependence is absorbed by the couplings through
the usual renormalization procedure. There remain contributions, however,
which contain inverse cut--off powers and depend on $T$ and $L_s$.

We want to argue now that the cut--off effects enter the free energy
density on the O($p^5$) level first.
At finite cut--off spatial rotation symmetry is lost, only the $90^\circ$
discrete symmetry of the square lattice remains. We need at least four
derivatives to observe the difference. Since every term contains at
least two $\vec{\pi}$ fields, the new term is O($p^5$) or higher.
Actually, there is a new contribution at O($p^5$) already.
The reduced symmetry allows the new four--derivative term
\bee
{1\over 2} g_4^{(4)}\sum_{i=1,2}
\left( \partial_i \partial_i {\bf S}
\partial_i \partial_i {\bf S} \right)
\,,
\label{33}
\ee
which is invariant under $90^\circ$ spatial rotation but not rotation
symmetric.
The corresponding leading term
\bee
{1\over 2} g_4^{(4)}\sum_{i=1,2}
\left( \partial_i \partial_i \vec{\pi}
\partial_i \partial_i \vec{\pi} \right)
\,,
\label{34}
\ee
is O($p^5$). The leading contribution to the free energy density
coming from this term is
\bee
\sim {T \over L_s^2} \sum_p{p_1^4+p_2^4 \over
p_0^2/c^2+p_1^2+p_2^2}\,,
\label{35}
\ee
where the denominator is the magnon propagator in leading order.
It is easy to show that eq.~(\ref{35}) has a non--zero,
O($p^5$) temperature and volume dependence.
Consider now the second source of cut--off dependence generated by the momentum
sums in chiral perturbation theory.
The leading O($p^3$) contribution is a $1$--loop graph corresponding
to a momentum sum.\footnote{The summand is essentially the logarithm of the
leading magnon propagator. See Section 5.}
The leading cut--off effect in bosonic sums and integrals is
$\sim(1/\Lambda^{cut})^2$. We expect therefore
$p^3 \cdot p^2/(\Lambda^{cut})^2 =$O($p^5$) cut--off dependent corrections.

These considerations show that the effective action in eq.~(\ref{30}), which
after restoring the dimensions has the form
\bee
\int_0^{\hbar/T}dt\int_0^{L_s}d^2x \, {1\over 2}\rho_s
\left[ {1\over c^2}\partial_{t}{\bf S}\partial_{t}{\bf S}+
\sum_{i=1,2}\partial_i{\bf S} \partial_i{\bf S} \right] \,,
\label{36}
\ee
is sufficient to derive the leading (O($p^3$)) and next--to--leading
(O($p^4$)) contributions to the free energy density.
These contributions depend therefore on two parameters $\rho_s$ and $c$
only and are cut--off independent.
Eq.~(\ref{36}) can be generalized trivially to O($N$) symmetry
by defining ${\bf S}$ as an $N$--vector and $\vec{\pi}$ as an $(N-1)$--vector.
In \cite{36}, the cut--off dependence has been investigated in the large--$N$
limit of the classical O($N$) ferromagnetic Heisenberg model.
The exact solution showed a cut--off dependence which is consistent with our
discussion above.

\bigskip
\section{\bf The free energy density $f(T,L_s)$ up to
next--to--leading order ("cubic" geometry)}

We take "$\hbar =c=1$" and restore the dimensions at the end.
It is useful to introduce the notations
\bee
L_t={1\over T},~~ V=L_s^2 \cdot L_t,~~ L=V^{1/3},~~x=(x_0=t,x_1,x_2)\,.
\label{37}
\ee
In the effective prescription the partition function is given by
\bee
Z=\prod_x \int d{\bf S}(x) \delta \left( {\bf S}^2(x)-1 \right) \cdot
\rme^{-{\cal A}({\bf S})}
\,,
\label{38}
\ee
where the action ${\cal A}$ is given in eq.~(\ref{30}).
As we discussed before, the field $\vec{\pi}(x)$ is small,
O($p^{1/2}$) which is the basis of the perturbative expansion in
eq.~(\ref{38}). The euclidean box $L_t \times L_s \times L_s$
is large compared to the scale of non--magnon excitations
and we shall assume that its shape is essentially cubic
\bee
L_s \gg {1\over \rho_s},~~~ L_t \gg {1\over \rho_s},~~~
l^3={L_t \over L_s}={\rm O}(1) \,.
\label{39}
\ee
Perturbation theory in a box of this kind has a technical problem
due to the existence of zero modes. The physical reason is that the
dominant ${\bf S}$--configurations have a net (staggered) magnetization,
i.e. a non--zero value for
\bee
{1\over V}\int d^3 x {\bf S}(x) \,,
\label{40}
\ee
which freely rotates around in the O($N$) group space. One has to separate
these modes before starting a perturbative expansion.
Since the technique is standard we quote the result only \cite{37,10}.
Eq.~(\ref{38}) can be written in the form
\beea
Z & = & \prod_x \left(\int d \vec{\pi}(x)
{1 \over \left(1-\vec{\pi}^2(x)\right)^{1/2}}\right)
\prod_{i=1}^{N-1}\delta\left({1\over V}\int d^3x\pi^i(x)\right) \times
\nonumber \\
& & \exp\left\{ -{\cal A}({\bf S})+
(N-1)\ln\left[{1\over V}\int d^3x \left( 1-\vec{\pi}^2(x)\right)^{1/2}\right]
\right\}
\,.
\label{41}
\eea
Going over to Fourier--space, we have
\bee
\prod_x \left(\int d \vec{\pi}(x) \right) \cdot
\prod_{i=1}^{N-1}\delta\left({1\over V}\int d^3x\pi^i(x)\right)
\rightarrow
\prod_k\left( \int d \vec{\tilde{\pi}_k} \right)
\prod_{i=1}^{N-1} \delta \left( \tilde{\pi}^i_{k=0} \right) \cdot
\rme^{ {N-1 \over 2} \ln V } \,.
\label{42}
\ee
Eliminating the zero modes therefore has the following consequences:
in the Fourier sums the $k=(0,0,0)$ mode should be left out, the action
receives an additional, field--dependent contribution (the second term
in the exponent in eq.~(\ref{41})) and a field independent, but size
dependent constant ${1\over 2}(N-1)\ln V$ occurs in the exponent.

One follows the rules of standard perturbation theory except that
in calculating Feynman graphs in the momentum sums the zero mode does not
enter. According to eqs.~(\ref{41},\ref{42}) in leading and
next--to--leading order the Lagrangean has the form
\beea
{\cal L}&=&{\cal L}_0+{\cal L}_1 \,, \nonumber \\
{\cal L}_0&=&{1\over 2}\rho_s \partial_{\mu}\vec{\pi}\partial_{\mu}\vec{\pi}
-{N-1 \over 2V} \ln V \,, \label{43} \\
{\cal L}_1&=&{1\over 2}\rho_s  \left( \vec{\pi}\partial_{\mu}\vec{\pi}\right)
\left( \vec{\pi}\partial_{\mu}\vec{\pi}\right) +
{N-1\over 2V} \vec{\pi}^2 \,. \nonumber
\eea
Here ${\cal L}_1$ is O($p^4$) and can be treated as a perturbation.
As discussed in Section~4, there are no $T$-- and $L_s$--dependent
cut--off effects up to the order of our calculation, O($p^4$).
Therefore, the results are universal, independent of the regularization
used in the effective theory. A conventional possibility is dimensional
regularization \cite{38}. In this case the factor $(1-\vec{\pi}^2(x))^{-1/2}$
in the measure in eq.~(\ref{41}) can be
suppressed \cite{39}.\footnote{Dimensional
regularization is very convenient, but certain aspects of it are rather formal.
We have checked that the final results of this section remain unchanged
if lattice regularization is used.}
One obtains for the free energy density
\beea
 & &    f=f_0+f_1 \nonumber \\
 & &   f_0= {N-1 \over 2 V} {{\sum}_k}'\ln k^2 -
         {N-1 \over 2 V} \ln V \,,  \label{44}\\
 & &   f_1=\langle {\cal L}_1 \rangle_0 \,, \nonumber
\eea
where ${\sum}'$ denotes a summation over the discrete momenta
$k_\mu=(k_0,k_1,k_2)$ of the periodic box $L_t\times L_s\times L_s$
with the $k_\mu=(0,0,0)$ mode left out, while $k^2=k_0^2+k_1^2+k_2^2$.
The notation $\langle ~~ \rangle_0$ refers to the expectation value
with the leading ${\cal L}_0$ in the exponent:
\bee
\langle {\cal L}_1 \rangle_0 ={1\over 2\rho_s}(N-1)\bar{G} (0)
\left. \partial_\mu^x \partial_\mu^y \bar{G} (x-y)\right|_{x=y}+
{1 \over 2\rho_s V} (N-1)^2  \bar{G} (0) \,,
\label{45}
\ee
where
\bee
\bar{G}(x)={1\over V} {{\sum}_k}'\,{\rme^{ikx} \over k^2} \,.
\label{46}
\ee
In calculating the momentum sums in eqs.~(\ref{44},\ref{45}) one
separates first the $V=\infty$ part, which is in general divergent.
This part is absorbed through the renormalization of the parameters
of the effective Lagrangean. In dimensional regularization this
renormalization is trivial, since the corresponding integrals are defined
to be zero.\footnote{In Sections~6 and 8, where lattice regularization will be
used, we carry through this renormalization process explicitly.}
The remaining finite, $T$-- and $L_s$--dependent part can be written
in a form which allows an easy numerical evaluation for any value
of the shape parameter $l^3=L_t/L_s$. Since these steps were repeatedly
discussed in the literature \cite{6,7,10}, we quote the final result only.
Eqs.~(\ref{44},\ref{45}) can be written as
\beea
& & f_0=q_1+{q_2\over V} -{N-1\over 2 V}
\left[ {1\over 3} \ln V + \beta_0 (l) \right] \,, \nonumber \\
& & f_1= -{(N-1)(N-2) \over 2 V}\cdot{1\over \rho_s L_s l}
\beta_1(l) \,, \label{47}
\eea
where $q_1$ and $q_2$ are two non--universal constants, $l$ and the functions
$\beta_n(l)$  were defined in eq.~(\ref{4}) and eqs.~(\ref{14a}--\ref{14}),
respectively.
By restoring the dimensions, eq.~(\ref{47}) leads to the result
quoted in eq.~(\ref{11}).

These results have been derived under the assumption that
$l=(\hbar c /TL_s)^{1/3}$ is O($1$). What happens if we take the $T \to 0$,
or $L_s \to \infty$ limit?

The $T\to 0$ ($L_t \to \infty$, or $l\to \infty$) limit corresponds to
an infinitely elongated cylinder geometry. As it is well known \cite{15,22,7}
in this geometry there are additional quasi--zero modes corresponding
to the slow rotation of the (staggered) magnetization of the $L_s\times L_s$
planes as we move along the $t$--direction. These quasi--zero modes
should be treated separately. The expansion used to derive eq.~(\ref{47})
breaks down in the $l\to\infty$ limit and does {\em not} give the correct
ground state energy.
For later reference we quote the result of this limit, nevertheless:
\beea
\left. {\rm eq.~(\ref{47})} \right|_{l\to\infty} : & &
f_0 \to  -{N-1\over 2} {\hbar c \over L_s^3} \cdot 1.437745 \,, \nonumber \\
 & & f_1 \to  {(N-1)(N-2) \over 24}
{(\hbar c)^2\over\rho_s L_s^4}  \,. \label{48}
\eea

There is an other way to see the problem with the $T\to 0$ limit.
In this limit we have effectively a $d=1$ non--linear $\sigma$--model.
In $d=1$, the $L_t \to \infty$ limit can not be interchanged
with the weak coupling expansion \cite{37}.
Since this problem is specific to $d=1$, we do not expect similar problems
in the $L_s\to \infty$ ($l\to 0$) limit, where we obtain a
two--dimensional slab.
Although there are new non--perturbative effects in this limit which
generate a finite mass $m$ to the magnons \cite{11,17},
this mass is exponentially
small in $1/T$. The effect of such a small mass is negligible in
the free energy density assuming that there are no infrared divergences
in the $m\to 0$ limit, i.e. assuming that the perturbative expansion
for $f$ in $d=2$ is infrared finite. This is actually the case \cite{41}.
Eq.~(\ref{47}) gives then
\bee
f(T)=q_1-(N-1){\zeta(3)\over 2\pi}T^3 +{\rm O}\left( T^5 \right) \,.
\label{49}
\ee
There is no $\sim T^4$ contribution, since
$\left. {f_1} \right|_{l\to 0} =0$.
This is the result (after restoring dimensions) announced in eq.~(\ref{16}).

\section{\bf Volume dependence of the ground state energy
and of the lowest lying excitations}

The special properties of the cylinder
geometry corresponding to this
problem have been observed and discussed before \cite{15,22,7}.
The next--to--leading volume correction to the ground state energy has
been calculated in $d=4$ for the symmetry group SU($n$)$\times$SU($n$)
relevant for the strong interactions \cite{7}.
The next--to--leading ground state energy for O($N$) and the volume
correction to the lowest excitation energies to be discussed here
are new results.

Since this problem is a somewhat non--trivial application of chiral
perturbation theory, we use this occasion to discuss some technical issues:
we consider a "physical" regularization (lattice, as opposed to dimensional
regularization of Section 5) and discuss the renormalization process
explicitly, especially the problem of power divergences. We give
also some details on the way collective coordinates are introduced in
treating the zero and quasi--zero modes.

We shall consider a $d$--dimensional euclidean cylinder
$L_s^{d-1}\times L_t$ in the $L_t \to \infty$ limit.
The AF Heisenberg model corresponds to $d=3$, $N=3$ in the following
expressions. We shall use a (hyper)cubic lattice to regularize the effective
theory. We put the lattice unit equal to $1$ ($a=1$) and denote the lattice
points by $x=(t,\vec{x})$, where $t=1,2,\ldots,L_t$, etc.
The Fourier transform is defined as
\bee
\pi(x)={1\over \sqrt{V}} \sum_k \rme^{ikx} \tilde{\pi}(k)\,,~~~
V=L_s^{d-1}\cdot L_t \,,
\label{50}
\ee
and we introduce the notation $\Delta_\mu$:
\bee
\Delta_\mu g(x) = g(x+\hat{\mu})-g(x) \,,
\ee
where $\hat{\mu}$ is the unit vector in the $\mu$--th direction.

\subsection{\bf Introducing collective coordinates}

Until $L_t=$O($L_s$), it is sufficient to take special care of the freely
moving total magnetization. For large  $L_t/L_s$, however, the magnetization
of distant time slices in the cylinder can differ significantly.
We have to treat these slowly moving variables (they are the
$k=(k_0,\vec{k}=\vec{0}$) modes) non--perturbatively.

The action in eq.~(\ref{38}) has the regularized form
\bee
{\cal A}({\bf S}) = {1\over 2} \rho_s^0 \sum_{x,\mu}
\Delta_\mu {\bf S}(x) \Delta_\mu {\bf S}(x) \,,
\label{52}
\ee
where we denoted the coupling by $\rho_s^0$ anticipating a non--trivial
renormalization. A possibility to introduce appropriate collective coordinates
is to insert into the path integral
\bee
1=\prod_t \left[ \int d\m (t) \prod_{n=0}^{N-1}
\delta \left( m^n(t)-{1\over V_s}\sum_{\vec{x}}S^n(t,\vec{x})\right) \right]
\,,
\label{53}
\ee
where $V_s=L_s^{d-1}$. We write
\beea
& & \m (t)=m(t) \e (t)\,, ~~~ \e(t)^2=1\,, \nonumber \\
& & d \m (t)=m(t)^{N-1}  dm(t) d\e (t)\,, \label{54} \\
& & \int d \e \equiv \kappa = {2 \pi^{N/2} \over \Gamma(N/2)} \,.
\nonumber
\eea
Introduce the O($N$) rotation $\Omega(t)$
\beea
& &\e (t)=\Omega(t) \n , ~~~ \n=(1,0,\ldots,0)\,, \nonumber \\
& & \Omega \in {\rm O}(N)\,, ~~~ \int d\Omega =1 \,.
\label{55}
\eea
Clearly, the vector $\e$ does not completely fix the matrix $\Omega$.
On the other hand, integrating over $\Omega$ rather than $\e$
gives an $\e$--independent constant factor only. The partition
function has the form
\beea
Z &=& \prod_x \int d{\bf S}(x)\delta \left( {\bf S}^2(x)-1 \right) \
\prod_t \int dm(t) m(t)^{N-1} \times  \nonumber \\
 & &\kappa \int d\Omega(t) \delta^{(N)}\left( m(t)\Omega(t)\n -
{1\over V_s}\sum_{\vec{x}}{\bf S}(t,\vec{x}) \right)
\exp\left\{ -{\cal A} \left( {\bf S} \right) \right\} \,.
\label{56}
\eea
Replace ${\bf S}$ by the new integration variable $\R$:
\bee
{\bf S}(t,\vx)=\Omega(t) \R(t,\vx) \,.
\label{57}
\ee
Since the measure is O($N$) invariant, we get
\beea
Z &=& \prod_x \int d\R(x)\delta \left( \R^2(x)-1 \right)
\prod_t \int dm(t) m(t)^{N-1} \times  \nonumber \\
 & &\kappa \int d\Omega (t)\delta^{(N)}\left( m(t)\n -
{1\over V_s}\sum_{\vec{x}}\R(t,\vec{x}) \right)
\exp\left\{ -{\cal A} \left( \Omega\R \right) \right\} \,.
\label{58}
\eea
Write
\bee
R^i=\pi^i\,,~~~i=1,\ldots,N-1\,,~~~ R^0=\left( 1-\vpi^2 \right)^{1/2} \,,
\label{59}
\ee
and integrate over the $\delta$--functions
\beea
& &Z = \prod_x \int d\vpi(x)
\prod_t  \int d\Omega (t) \prod_{i=1}^{N-1}
\delta \left( {1\over V_s} \sum_{\vx} \pi^i (t,\vx) \right) \cdot
 \exp \left\{ -{\cal A}_1(\Omega,\vpi)  \right\}  \,, \nonumber \\
& & {\cal A}_1 (\Omega,\vpi)= {\cal A} (\Omega\R)+
{1\over 2} \sum_x \ln \left( 1-\vpi^2(x) \right) -  \label{60} \\
& & ~~~~~~~~~~~~
(N-1)\sum_t \ln \left( {1\over V_s} \sum_{\vx} R^0(t,\vx)\right)
 - L_t \ln \kappa\,.  \nonumber
\eea
Going over to Fourier variables, eq.~(\ref{60}) leads to
\beea
& & Z  = \left( \prod_k \prod_{i=1}^{N-1}\int d\tpi^i(k) \right)
\left( \prod_t  \int d\Omega (t) \right)
\left( \prod_{k^0} \prod_{i=1}^{N-1}
\delta \left( \tpi^i (k^0,\vec{k}=\vec{0}) \right)  \right) \times
 \nonumber \\
& &  ~~~~ \exp \left\{ -{\cal A}_1(\Omega,\vpi)
 -L_t {N-1 \over 2} \ln V_s  \right\} \,.
\label{61}
\eea

\subsection{\bf Integrating over the collective variables}

The action ${\cal A}(\R)$ can be written as
\bee
{\cal A}(\R) ={1\over 2} \rho_s^0 \sum_{x,\mu}
\left( \R (x+\hat{\mu})-\R(x) \right)^2 =
dV \rho_s^0 - \rho_s^0 \sum_{x,\mu}\R (x+\hat{\mu})\cdot\R(x) \,.
\label{62}
\ee
Since $\Omega$ is independent of $\vx$, we have
\bee
{\cal A}(\Omega \R) = dV \rho_s^0- \rho_s^0 \sum_x \sum_{i=1}^{d-1}
\R (x+\hat{i})\R(x) - \rho_s^0 \sum_{t,\vx}
\Omega(t+1)\R(t+1,\vx)\cdot\Omega(t)\R(t,\vx) \,.
\label{63}
\ee
Consider the sum in the last term
of eq.~(\ref{63})
\bee
\sum_{\vx} \left[ \Omega(1)^T \Omega(2) \R(2,\vx)\cdot \R(1,\vx)
+ \Omega(2)^T \Omega(3) \R(3,\vx)\cdot \R(2,\vx) + \ldots \right] \,.
\label{64}
\ee
As eq.~(\ref{55}) shows, the direction $\e(t)$ of the magnetization
of a time slice is not influenced by the part of $\Omega(t)$
which leaves the vector $\n$ invariant. These are spurious degrees
of freedom in the O($N$) matrix $\Omega$. We shall get rid of these
spurious variables by changing integration variables in eq.~(\ref{61})
appropriately.
Replace the integration variable $\Omega(2)$ in eq.~(\ref{61}) by
\bee
V(2)=\Omega(1)^T \Omega(2) \,, \label{65A}
\ee
and introduce the rotated vectors
\bee
\hat{\R}(2,\vx)= \hat{V}(2) \R(2,\vx)\,.
\label{65}
\ee
Here $\hat{V}(2)$ is an O($N$) element with the structure
\bee
\hat{V}(2)=\left( \begin{array}{cccc}
                      1 & 0 & \cdots & 0 \\
                      0 & \ddots  &        &   \\
                \vdots  &   & \hat{V}_{ik}(2) &  \\
                      0 &   &        & \ddots  \\
                   \end{array} \right) \,,
{}~~~~i,k=1,2,\ldots,N-1 \,.
\label{66}
\ee
The O($N-1$) rotation $\hat{V}_{ik}(2)$ serves to exclude the spurious
degrees of freedom in $V(2)$ and will be specified further later.
Eq.~(\ref{65}) is an O($N-1$) rotation on the integration variables
$\pi^i(2,\vx)$
\bee
\hat{\pi}^i(2,\vx)=\hat{V}_{ik}(2)\pi^k(2,\vx) \,.
\label{67}
\ee
We can introduce $\hat{\pi}^i(2,\vx)$ as new integration variables
in the $t=2$ plane in eq.~(\ref{61}). The measure is invariant and
the $\delta\left(\tpi^i(k_0,\vec{k}=\vec{0})\right)$ constraint is as well
since the rotation is independent of $\vx$.
The terms in the exponent in eq.~(\ref{61}) are also invariant, except
that eq.~(\ref{64}) goes over to
\bee
\sum_{\vx} \left[ V(2)\hat{V}(2)^T \hat{\R}(2,\vx) \cdot\R(1,\vx)+
\left(\Omega(1)V(2)\hat{V}(2)^T \right)^T \Omega(3)\R(3,\vx)\cdot
\hat{\R}(2,\vx) \ldots \right]\,.
\label{68}
\ee
Now we can repeat this procedure with the second term in eq.~(\ref{68}).
We replace the integration variable $\Omega(3)$ by
\bee
V(3)=\left( \Omega(1)V(2)\hat{V}(2)^T \right)^T \Omega(3) \,,
\label{69}
\ee
and transform the variables in the $t=3$ plane by the $\vx$--independent
rotation
\bee
\hat{\R}(3,\vx)=\hat{V}(3)\R(3,\vx) \,,
\label{70}
\ee
where $\hat{V}(3)$ has the structure of eq.~(\ref{66}). Repeating these
steps in the $t=4,\ldots$ planes, eq.~(\ref{68}) goes over
\bee
\sum_{\vx} \left[ V(2)\hat{V}(2)^T \hat{\R}(2,\vx) \cdot \R(1,\vx)+
V(3)\hat{V}(3)^T \hat{\R}(3,\vx) \cdot \hat{\R}(2,\vx) +
\ldots \right] \,.
\label{71}
\ee
Boundary effects can be neglected for $L_t\to \infty$.
Introducing the notation
\bee
\tV (t-1)= V(t)\hV(t)^T
\label{72}
\ee
and suppressing the "hat" on $\hat{\R}$, we get
\beea
& &Z = \left( \prod_k \int d\tpi^i(k) \right)
\left( \prod_t  \int dV(t) \right)
\left( \prod_{k^0} \prod_{i=1}^{N-1}
\delta \left( \tpi^i (k^0,\vec{k}=\vec{0})\right)  \right)\times\nonumber \\
& & \exp -\left\{  {\cal A}(V,\R)
+ {1\over 2} \sum_x \ln \left( 1-\vpi^2(x) \right) - \right. \label{73} \\
& & ~~~~~~~~~ \left.
(N-1)\sum_t \ln \left( {1\over V_s} \sum_{\vx} R^0(t,\vx)\right)
- L_t \ln \kappa -L_t {N-1 \over 2} \ln V_s \right\} \,, \nonumber
\eea
where
\beea
{\cal A}(V,\R) & = & d V \rho_s^0 - \rho_s^0 \sum_x \sum_{i=1}^{d-1}
\R(x+\hat{i})\cdot \R(x)-
\rho_s^0\sum_{t,\vx} \tV (t)\R(t+1,\vx)
\cdot \R(t,\vx) \nonumber \\
 & = & {1\over 2} \rho_s^0 \sum_{x,\mu} \Delta_{\mu}\R \Delta_{\mu}\R
- \rho_s^0 \sum_{t,\vx} Q(t) \R(t+1,\vx) \cdot \R(t,\vx) \,,
\label{74}
\eea
with the notation
\bee
Q(t)=\tV (t)-1 \,.
\label{75}
\ee
We write the exponent in eq.~(\ref{73}) as
\bee
\{ \ldots \} = {\cal A}^{(\pi)} + {\cal A}^{(\pi,V)} -L_t \ln \kappa
-L_t {N-1 \over 2}
\ln V_s \,,
\label{76}
\ee
where ${\cal A}^{(\pi)}$ is the $\pi$--dependent part, while
${\cal A}^{(\pi,V)}$ is the $V$--dependent part including the $\pi$--$V$
interactions. Since we have replaced the dangerous components
$\tilde{\pi}^i(k_0,\vec{k}=\vec{0})$ by collective variables, we
can consider the remaining $\tilde{\pi}$ variables as small, and expand.
Eqs.~(\ref{73},\ref{74}) give for the $\pi$--dependent part
\beea
{\cal A}^{(\pi)}&=& {\cal A}^{(\pi)}_0 +{\cal A}^{(\pi)}_1 +\ldots \,,
\nonumber \\
{\cal A}^{(\pi)}_0 &=& {1\over 2}\rho_s^0 \sum_{x,\mu}
\Delta_{\mu}\vpi \Delta_{\mu}\vpi \,, \label{77} \\
{\cal A}^{(\pi)}_1 &=&
{1\over 2}\rho_s^0 \sum_{x,\mu}
\left[ (\vpi \Delta_{\mu}\vpi )(\vpi \Delta_{\mu}\vpi ) +
(\vpi \Delta_{\mu}\vpi )(\Delta_{\mu}\vpi \Delta_{\mu}\vpi )+
\phantom{{1\over 4}} \right.
\nonumber \\
& & \left. {1\over 4}(\Delta_{\mu}\vpi \Delta_{\mu}\vpi )
(\Delta_{\mu}\vpi \Delta_{\mu}\vpi ) \right]
+{N-1 \over 2V_s} \sum_x \vpi^2
-{1\over 2} \sum_x \vpi^2 (x) \,. \nonumber
\eea
The last term in ${\cal A}^{(\pi)}_1$ comes from the measure, it plays
the role of a counterterm in the renormalization as we shall see.
In deriving eq.~(\ref{77}) we used the identity
\bee
\Delta_{\mu} \left( f(x)g(x) \right) = \Delta_{\mu} f(x) \cdot g(x)
+f(x) \cdot \Delta_{\mu} g(x) + \Delta_{\mu} f(x) \cdot \Delta_{\mu} g(x)\,,
\label{78}
\ee
(no summation over $\mu$).
The $V$--dependent part is generated by the last term in eq.~(\ref{74}).
Expanding in $\pi$, we get for this term
\bee
-\sum_t \left\{ Q_{00}(t)\rho_s^0 V_s
 -  Q_{00}(t) A(t) + Q_{ik}(t) B_{ik}(t)+   Q_{ik}(t) C_{ik}(t) +
{\rm O}(\pi\pi\pi) \right\} \,,
\label{79}
\ee
where
\beea
A(t) & = & \rho_s^0 \sum_{\vx} \left[ \vpi^2 (t,\vx) +
{1\over 2} \Delta_0 \left(\vpi^2(t,\vx) \right)
\right] \,, \label{80} \\
B_{ik}(t) & = & \rho_s^0 \sum_{\vx} \left[ \vpi_i(t,\vx)\vpi_k(t,\vx) +
{1\over 2} \Delta_0 \vpi_i(t,\vx)\vpi_k(t,\vx) +
{1\over 2}\vpi_i(t,\vx) \Delta_0 \vpi_k(t,\vx) \right] \,, \nonumber \\
C_{ik}(t) & = & {1\over 2}\rho_s^0 \sum_{\vx} \left[
 \Delta_0 \vpi_i(t,\vx)\vpi_k(t,\vx) -
\vpi_i(t,\vx) \Delta_0 \vpi_k(t,\vx)
\right] \,. \nonumber
\eea
As eqs.~(\ref{73},\ref{79}) show, in every time slice $t$ we have to
perform the group integral
\bee
\int dV \exp \left\{
\rho_s^0 L_s^{d-1} \left( \tV_{00} -1 \right)
-Q_{00}A + Q_{ik}B_{ik} +
 Q_{ik}C_{ik}  \right\} \,,
\label{81}
\ee
where $\tV $ and $Q$ are related to $V$
 through eqs.~(\ref{72},\ref{75}).
Due to the form of $\hat{V}$, eq.~(\ref{66}), we have
\bee
\tV_{00}=V_{00}\,, ~~~Q_{00}=V_{00}-1 \,.
\label{82}
\ee
For $\rho_s^0 L_s^{d-1}$ large, the typical $V$ matrices contributing to
the integral in eq.~(\ref{81}) are
\bee
\tV_{00}=V_{00}=1-\epsilon^2\,,~~~ \epsilon^2={\rm O}\left(
{1 \over \rho_s^0 L_s^{d-1} } \right)\,,
\label{83}
\ee
where $\epsilon$ is small. Then $V_{i0}=$O($\epsilon$).
We shall now specify the matrix $\hat{V}$ in eq.~(\ref{66}) in terms of
the matrix elements $V_{ik}$ in such a way that the integrand in
eq.~(\ref{81}) becomes free of the spurious degrees of freedom discussed
before. We shall take $\hat{V}_{ik}$ essentially equal to $V_{ik}$, plus
corrections needed to make $\hat{V}_{ik}$ an element of O($N-1$):
\beea
& & \hat{V}_{ik}=\alpha_i V_{ik} + \sum_j \beta_{ij} V_{jk}\,,~~
{\rm no~ sum~ over}~i\,, \nonumber \\
& & \beta_{ij}=\beta_{ji}\,,~~~ \beta_{ii}=0 \,.
\label{84}
\eea
For small $\epsilon$ we have
\beea
& & \alpha_i = 1+{1\over 2}\left( V_{i0} \right)^2 +{\rm O}(\epsilon^3) \,,
 \label{85} \\
& & \beta_{ij} ={1\over 2} V_{i0} V_{j0}  +{\rm O}(\epsilon^3)\,,
{}~~~i\neq j \,. \nonumber
\eea
Then $\tV$ in eq.~(\ref{72}) has the form
\bee
\tV = V \hat{V}^T=
       \left( \begin{array}{ccccc}
                      V_{00} &  & -V_{00}V_{i0}  & &\cdots  \\
                       & \ddots &   &  & \\
                             &  &        &   \\
                V_{i0}       & & 1-{1\over 2}V_{i0}^2 &
                                               -{1\over 2}V_{i0}V_{k0} & \\
     \phantom{\vdots}  &  & -{1\over 2}V_{i0}V_{k0} &  & \\
                  \vdots      &  &   &  &  \ddots\\
                   \end{array} \right)
{}~~~~ +{\rm O}(\epsilon^3) \,.
\label{86}
\ee
Only $V_{00}$ and $V_{i0}$, with $V_{00}^2 + \sum_i V_{i0}^2 =1$ enter in
eq.~(\ref{86}), representing the motion of $\e$ in eq.~(\ref{55}) without
additional spurious degrees of freedom.
Eq.~(\ref{86}) shows that $Q_{00} \sim Q_{ik} \sim$O($1/\rho_s^0L_s^{d-1}$).
This is just the order we want to go. Expanding eq.~(\ref{81}) in terms
of $Q_{00}$ and $Q_{ik}$, we get
\bee
\int dV \exp \left\{ \rho_s^0 L_s^{d-1} Q_{00} \right\}
\left[ 1 -Q_{00}A +\sum_{i,k} Q_{ik}B_{i,k} \right] \,,
\label{87}
\ee
where
\bee
Q_{00}=V_{00} -1 \,,~~~Q_{ik}=-{1\over 2}V_{i0} V_{k0} \,.
\label{88}
\ee
 In eq.~(\ref{87}) we can replace $A$ and $B_{ik}$ by their expectation
value calculated with the leading part of $A^{(\pi)}$
(the first term in eq.~(\ref{77})):
\beea
& & A \to \langle A \rangle = (N-1) V_s D^*(0) \,, \label{89} \\
& & B_{ik} \to \delta_{ik} \langle B \rangle = \delta_{ik}V_s \left( D^*(0)
+\Delta_0 D^*(0) \right) \,, \nonumber
\eea
where
\bee
D^*(0) = {1\over V} \sum_k {}^* {1 \over \sum_{\mu} 4 \sin^2 (k_{\mu}/ 2)}
\,.
\label{90}
\ee
In eq.~(\ref{90}) a constrained summation over the momenta occurs,
\bee
\sum_k {}^* \equiv \sum_{k \neq (k^0,\vec{k}=\vec{0})} \,.
\label{91}
\ee
Using
\bee
\sum_{i=1}^{N-1} Q_{ii} = -{1\over 2} \sum_{i=1}^{N-1} V_{i0}^2 =
-{1\over 2} \left( 1 -V_{00}^2 \right) = Q_{00} +{\rm O}(\epsilon^3)\,,
\label{92}
\ee
we can write eq.~(\ref{87}) in the form
\bee
\int dV(t) \exp \left\{ \Theta_{eff} Q_{00} \right\}\,,
{}~~~\Theta_{eff}=\rho_s^0 L_s^{d-1} - \langle A \rangle +
\langle B \rangle \,.
\label{92A}
\ee
At this stage it is clear that eq.~(\ref{92A}) represents the motion of
an O($N$) rotator with inertia $\Theta_{eff}$.
Indeed, we have
\beea
& & Q_{00}(t)=V_{00}(t)-1 = \left( \Omega(t-1)^T \Omega(t) \right)_{00} -1
=\Omega(t-1) \n \cdot \Omega(t) \n -1 \nonumber \\
& & ~~~~~~~~~~~ = -{1\over 2} \left( \e(t) -\e(t-1) \right)^2 \,,
\label{92B}
\eea
and the integral in eq.~(\ref{92A}) can be written as
\bee
{1\over \kappa} \int d \e(t) \exp \left\{
 -{1\over 2} \Theta_{eff} \left( \e(t) -\e(t-1) \right)^2 \right\} \,.
\label{92C}
\ee
Here we used that $dV(t)=d\Omega(t)= d\e(t)/\kappa$.

The group integral in eq.~(\ref{92A}) can be easily performed.
For large $\rho_s^0 V_s$ we get
\bee
\omega \exp \left\{ -{N-1 \over 2} \ln \left( \rho_s^0 V_s \right)
-{(N-1)(N-3) \over 8\rho_s^0 V_s}+{N-1 \over 2\rho_s^0 V_s }
\left( \langle A \rangle - \langle B \rangle \right)
\right\}
\,,
\label{93}
\ee
where
\bee
\omega=\Gamma \left({N\over 2}\right) {1\over 2 \sqrt{\pi}}2^{N-1\over 2} \,.
\label{94}
\ee
The corresponding part of the free energy density reads
\beea
 f^{(\pi,V)} &=& {N-1\over 2 V_s} \ln \left(\rho_s^0 V_s\right) +
{(N-1)(N-3)\over 8 \rho_s^0 V_s^2} - \nonumber \\
& & {N-1\over 2 \rho_s^0 V_s}
\left[ (N-2)D^*(0)-\Delta_0D^*(0)\right] -{1\over V_s}\ln \omega \,.
\label{95}
\eea

\subsection{\bf Contribution from the perturbative fast modes}

The leading contribution is given by the first term in eq.~(\ref{77})
which leads to
\bee
f_0^{(\pi)}= {N-1 \over 2V}{\sum_k}^* \ln \left( {1\over 2\pi}
\rho_s^0 d(k) \right)\,,
\label{96}
\ee
where
\bee
d(k)=4\sum_{\mu} \sin^2{k_{\mu}\over 2}
\label{97}
\ee
and ${\sum}^*$ is defined in eq.~(\ref{91}). We can write
\bee
{\sum_k}^* \ln \left( d(k) \right) =
\sum_{k\neq(0,\ldots,0)} \ln \left( d(k) \right) -
\sum_{k^0 \neq 0} \ln \left( d(k^0,\vec{k}=0) \right) \,.
\label{98}
\ee
The sums on the right hand side of eq.~(\ref{98}) are standard in
chiral perturbation theory and can be evaluated easily.
In the $L_t\to\infty$ limit one obtains\footnote{We suppress the non--universal
terms $\sim$const. or $\sim$const./$V$ in $f$.}
\bee
f_0^{(\pi)} =-{N-1\over 2 L_s^d}
\left[ \alpha_{-1/2}^{(d-1)}(1)+2-{2\over d} \right]
-{N-1 \over 2 L_s^{d-1}} \ln{\rho_s^0\over 2\pi} \,,
\label{99}
\ee
where $\alpha_p^{(d)}$ are the "shape coefficients" defined earlier for
$d=3$ (eqs.~(\ref{12}--\ref{14})). In eq.~(\ref{99}) we need the shape
coefficients of a $(d-1)$ dimensional cube only which are listed
in Table~1 in ref.~\cite{10}.
The next--to--leading correction is given by the expectation value of
${\cal A}_1^{(\pi)}$ in eq.~(\ref{77}). Using the relations (Appendix A)
\beea
& & \left. \Delta_{\mu}^x \Delta_{\mu}^y D^*(x-y) \right|_{x=y}
= - 2 \Delta_{\mu} D^*(0) ~~~({\rm no~sum~over~} \mu)\,, \nonumber \\
& & 2\sum_{\mu}\Delta_{\mu} D^*(0)={1\over V_s} -1 \,, \label{100}
\eea
one obtains
\bee
f_1^{(\pi)}={1\over \rho_s^0}\left[ -{N-1\over 2}
\Delta_{\mu} D^*(0) \Delta_{\mu} D^*(0)  +
{(N-1)(N-2)\over 2V_s}D^*(0) \right] \,.
\label{101}
\ee

\subsection{\bf The ground state energy}

Collecting the size dependent constants
(the last two terms in eq.~(\ref{76})),
the contributions from the collective coordinates (eq.~(\ref{95})) and
from the fast modes (eq.~(\ref{101})) we get a surprisingly simple
result for the ground state density
\beea
\epsilon_0(L_s) & = & \lim_{T\to 0} f(T,L_s) \nonumber \\
 & = & - {N-1\over 2 L_s^d} \left[
\alpha_{-1/2}^{(d-1)}(1) +2-{2\over d } \right]
\label{102} \\
 &  & - {N-1\over 2\rho_s^0} \left[ \Delta_{\mu} D^*(0) \Delta_{\mu} D^*(0)
-{1\over V_s} \Delta_0 D^*(0) \right] +
{(N-1)(N-3)\over 8\rho_s^0 V_s^2} \,. \nonumber
\eea
In obtaining eq.~(\ref{102}) we used the definitions of $\kappa$ and
$\omega$ in eq.~(\ref{54}) and (\ref{94}), respectively.
In order to proceed further we need the explicit form of the lattice
Green functions in eq.~(\ref{102}). As it is discussed in Appendix A,
all the dangerous cut--off dependence cancels in eq.~(\ref{102}),
and up to a non--universal volume independent constant we can write
\bee
\Delta_{\mu} D^*(0) \Delta_{\mu} D^*(0) -{1\over V_s}\Delta_0 D^*(0)
\to - {1\over 4 V_s^2} \,.
\label{103}
\ee

The spin--stiffness occurs in eq.~(\ref{102}) first in the correction
only, we can therefore replace $\rho_s^0$ by the renormalized, physical
$\rho_s$. (The renormalization properties will be discussed in 6.7).
We get then $(V_s=L_s^{d-1})$
\bee
\epsilon_0(L_s)-\epsilon_0(\infty)=-{N-1\over 2 L_s^d}
\left[ \alpha_{-1/2}^{(d-1)}(1)+2-{2\over d} \right]
+{(N-1)(N-2)\over 8\rho_s V_s^2} \,.
\label{104}
\ee
The shape coefficients $\alpha_{-1/2}^{(d-1)}(1)$ are listed
in Table~1 in ref~\cite{10}. The result quoted in eq.~(\ref{18})
is obtained after restoring the dimensions and setting $d=3$.
For $N=4$, $d=4$ eq.~(\ref{104}) agrees with the result in ref.~\cite{7}
obtained for SU($2$)$\times$SU($2$).

\subsection{\bf An alternative, intuitive method}

Since the previous calculation is rather involved, it is
reassuring that the result in eq.~(\ref{104}) can be reproduced in
a simple intuitive way. The following considerations offer additional insight
into the physics of the problem also.

In Section 5 we studied the free energy density $f(T,L_s)$ under conditions
$T \ll \rho_s$, $1/L_s \ll \rho_s$, $TL_s=$fixed constant.
Very low temperature corresponds chosing this fixed constant very small,
but even in this limit the temperature $T$ remains $\sim 1/L_s$.
Standard Goldstone boson (magnon) excitations carry momenta
$\sim 1/L_s$ and they freeze out at these low temperatures.
The rotator excitations, however, carry energies $\sim 1/L_s^2$ and for those
excitations the temperature is high even in this limit. They are fully
excited and included in the free energy. This is the reason why eq.~(\ref{48})
disagrees with the correct ground state energy density eq.~(\ref{18}).
Rather, eq.~(\ref{48}) should agree with the free energy density coming
from the O($N$) rotator
\bee
\hat{H}={1\over 2 \Theta_{eff}} \hat{\bf L}^2 +E_0(L_s)
\label{E1}
\ee
in the {\em high} temperature limit, where
$\Theta_{eff} \approx\rho_s L_s^2$ and
$E_0(L_s)$ is the ground state energy we are looking for.
The operator $\hat{\bf L}^2$ is, in coordinate representation, the Laplace
operator on the $N-1$ dimensional unit sphere.  This matching condition
can be used to calculate the ground state energy $E_0(L_s)$.
We need the high temperature expansion of the partition function
\bee
Z^{rot} = {\rm Tr} \exp \left( -{\hat{H}\over T}\right) =
\exp \left( -{L_s^2 \epsilon_0(L_s) \over T} \right)
{\rm Tr} \,\exp \left( -{ \hat{\bf L}^2\over  2 \Theta_{eff} T}\right) \,.
\label{E2}
\ee
An elegant way to obtain the result is to use Seeley's expansion
giving \cite{42}
\beea
\left. {\rm Tr} \, \exp \left( -{ \hat{\bf L}^2 t}\right)
\right|_{t\to 0} &=& A \,t^{-(N-1)/2}
\left[1+{(N-1)(N-2)\over 6}\, t  \right. \nonumber \\
 & & \left. +{(N-1)(N-2)(5N^2-17N+18)\over 360} \, t^2 \ldots \right] \,,
\label{E3}
\eea
with the normalization factor $A=\pi^{1/2} 2^{-N+2}/ \Gamma(N/2)$, which
will not be important for us.
The free energy density predicted by the rotator problem has the form
\bee
f^{rot}=\epsilon_0(L_s)-{N-1\over 2}{T\over V_s} \ln \left( T V_s \right)
-{(N-1)(N-2)\over 12 \rho_s V_s^2} + \ldots \,.
\label{E4}
\ee
Matching eq.~(\ref{E4}) with eq.~(\ref{48}) we find the next--to--leading
correction to $\epsilon_0(L_s)$
\bee
{(N-1)(N-2)\over 12 \rho_s V_s^2}+{(N-1)(N-2)\over 24 \rho_s V_s^2}=
{(N-1)(N-2)\over 8 \rho_s V_s^2}\,,
\label{E5}
\ee
which is the result obtained before, eq.~(\ref{104}).

\subsection{\bf The lowest lying excitations and their volume dependence}

The collective slow modes correspond to an O($N$) rotator with inertia
$\Theta \sim L_s^{d-1}$. The corresponding excitations are $\sim 1/L_s^{d-1}$,
lying much below the excitations of the fast $\pi$--modes which are
$\sim 1/L_s$. Using eqs.~(\ref{88},\ref{89},\ref{92}) we can write
eq.~(\ref{87}) in the form
\bee
\int dV \exp \left\{ \left(\rho_s^0 L_s^{d-1}-\langle A \rangle +
\langle B \rangle \right) \left( V_{00} -1 \right) \right\} \,.
\label{105}
\ee
Eq.~(\ref{105}) shows that the $\pi$--modes modify the inertia of the O($N$)
rotator to
\bee
\rho_s^0 L_s^{d-1} \to \rho_s^0 L_s^{d-1}-\langle A \rangle +
\langle B \rangle  = V_s \left[ \rho_s^0 -(N-2)D^*(0)+\Delta_0 D^*(0) \right]
\,.
\label{106}
\ee
As we shall discuss in 6.7, eq.~(\ref{122}), the relation between
the bare and renormalized spin stiffness in our order reads
\bee
\rho_s^0=\rho_s+(N-2)D_{\infty}(0)+{1\over 2d} \,,
\label{107}
\ee
where $D_{\infty}(x)$ is the infinite volume lattice propagator.
Using the explicit form of the Green functions (Appendix A) we get
\bee
\Theta_{eff} = \rho_s L_s^{d-1}
\left[ 1 + {N-2 \over 4\pi \rho_s L_s^{d-2}}
\left( - \alpha_{1/2}^{(d-1)}(1) +2 {d-1\over d-2} \right) \right]
\,.
\label{108}
\ee
The excitations of the O($N$) rotator are given by
\bee
E_j-E_0={j(j+N-2) \over 2 \Theta_{eff}} \,,
\label{109}
\ee
which, at $d=3$, leads to the result quoted in eq.~(\ref{19}).

\subsection{\bf Renormalizations. Relations at infinite volume and zero
temperature}

The physical parameters $\rho_s$, $c$ and $\Ms$ are defined at infinite
spatial volume and zero temperature. In the following we study briefly
their renormalization with lattice regularization in $d=3$ and, as a side
product, we rederive the results quoted in eq.~(\ref{9}).

Switch on an external symmetry breaking staggered field
${\bf H}_s=(H_s,0,\ldots,0)$. The corresponding leading term in the
effective action reads \cite{10}
\bee
-\sum_x\Ms^0 {\bf H}_s{\bf S}(x) \,,
\label{110}
\ee
where $\Ms^0$ is the bare staggered magnetization. We consider a $d(=3)$
dimensional infinitely large hypercubic lattice. There are no special
zero modes in this case. Expanding the action (eq.~(\ref{52}) and
eq.~(\ref{110})) and the measure in the $\vpi$ fields one obtains
\bee
Z=\prod_x \int d \vpi(x) \exp\left\{-\sum_x({\cal L}_0+{\cal L}_1 +\ldots)
\right\} \,,
\label{111}
\ee
where
\beea
{\cal L}_0 &=& {1\over 2}\rho_s^0 \left( \sum_{\mu} \Delta_{\mu}\vpi
\Delta_{\mu}\vpi + M_0^2 \vpi^2 \right) - \Ms^0 H_s\,, \nonumber \\
{\cal L}_1 &=& {1\over 2}\rho_s^0 \left\{ \sum_{\mu}
\left[ (\vpi \Delta_{\mu}\vpi)(\vpi \Delta_{\mu}\vpi)+
(\vpi \Delta_{\mu}\vpi)(\Delta_{\mu}\vpi \Delta_{\mu}\vpi)+
{1\over 4}(\Delta_{\mu}\vpi \Delta_{\mu}\vpi)
  (\Delta_{\mu}\vpi \Delta_{\mu}\vpi) \right.
\right] \nonumber \\
& & \left. +{1\over 4}M_0^2 \left(\vpi^2 \right)^2 \right\}
-{1\over 2}\vpi^2 \,, \label{112}
\eea
with the notation
\bee
M_0^2={\Ms^0 H_s \over \rho_s^0} \,. \label{113}
\ee
Let us calculate first the (staggered) magnetization:
\bee
\Ms^H ={1\over V}{1\over Z}{\partial \over \partial H_s}Z =
\Ms^0 \left\langle \left(1-\vpi^2(0)\right)^{1/2} \right\rangle =
\Ms^0\left(1-{N-1\over 2 \rho_s^0}D_{\infty}^{(M_0)}(0) \right) \,.
\label{114}
\ee
Eq.~(\ref{114}) and the subsequent equations of this section are understood
as relations in the first non--trivial order. In eq.~(\ref{114})
$D_{\infty}^{(M_0)}(x)$ is the infinite volume massive lattice
propagator
\bee
D_{\infty}^{(M_0)}(x) = \int_{-\pi}^{\pi} {d^dk \over (2\pi)^d}
{\rme^{ikx}\over d(k)+M_0^2}\,, \label{115}
\ee
where $d(k)$ is defined in eq.~(\ref{97}). At $H_s=0$, $\Ms^H$ goes over
to the spontaneous magnetization $\Ms$ which determines the relation
between the bare and renormalized magnetizations
\bee
\Ms=\Ms^0 \left(1-{N-1\over 2 \rho_s} D_{\infty}^{(0)}(0)
\right) \,, \label{116}
\ee
where we replaced $\rho_s^0$ by $\rho_s$ since the difference is of higher
order. Eqs.~(\ref{114},\ref{116}) and the relation
\bee
D_{\infty}^{(M_0)}(0)=D_{\infty}^{(0)}(0)-{M_0\over 4\pi}\,,~~~(d=3)
\label{117}
\ee
lead to the result
\bee
\Ms^H=\Ms \left(1+{N-1\over 8\pi}\cdot{M\over\rho_s}\right) \,, \label{118}
\ee
where we replaced $M_0$ by $M=\Ms H_s /\rho_s$.
After restoring the dimensions, eq.~(\ref{118}) gives
the result quoted in eq.~(\ref{9}). In eq.~(\ref{116}) $D_{\infty}^{(0)}(0)$
is linear in the cut--off in $d=3$. This power divergence is absorbed in the
bare parameter $\Ms^0$ leading to a cut--off independent, universal
relation eq.~(\ref{118}).

The procedure is very similar in case of the spin stiffness.
A possibility to study the spin stiffness is to calculate the two--point
function of Goldstone bosons. This Green--function should have a pole
at $k^2=0$ (for $H_s=0$) which fixes the wave function renormalization
factor $z$. For the two--point function one obtains
\bee
\left. \sum_x \rme^{-ikx}{1\over Z}{\delta \over \delta H_s^i(x)}
{\delta \over \delta H_s^j(0)} Z \right|_{\stackrel{H_s\to 0}{k\to 0}}
={\delta_{ij}\over k^2}{\left(\Ms^0\right)^2\over \rho_s^0}
\left[ 1-{1\over \rho_s^0}\left(D_{\infty}^{(0)}(0)
-{1\over 2d}\right)\right] \,.
\label{119}
\ee
The residue of the pole gives $z$:
\bee
z={\left( \Ms^0\right)^2\over \rho_s^0}
\left[ 1-{1\over \rho_s^0}\left(D_{\infty}^{(0)}(0)
-{1\over 2d}\right)\right] \,.
\label{120}
\ee
Using the Ward identity \cite{43,10}
\bee
\rho_s={\Ms \over z^{1/2}}\,, \label{121}
\ee
eqs.~(\ref{116},\ref{120}) determine the relation between $\rho_s$ and
$\rho_s^0$:
\bee
\rho_s=\rho_s^0 \left[ 1-{1\over\rho_s}\left( (N-2)D_{\infty}^{(0)}(0)
+{1\over 2d} \right) \right] \,,~~~d=3\,. \label{122}
\ee
Repeating the calculation for the two--point function in eq.~(\ref{119})
with $H_s \neq 0$, the pole position is shifted from $k^2=0$ which determines
the Goldstone boson mass. One obtains
\bee
\left( M^H \right)^2 = M_0^2 \left[ 1+{1\over 2\rho_s^0}
\left( (N-3)D_{\infty}^{(M_0)}(0) +{1\over d}\right) \right] \,. \label{123}
\ee
In the limit $H_s \to 0$, eqs.~(\ref{116},\ref{122},\ref{123}) are consistent
with the Ward identity \cite{10}
\bee
\left. \left( M^H \right)^2 \right|_{H_s\to 0} ={\Ms\over \rho_s} H_s \,.
\label{124}
\ee
The residue of the pole $\left( k^2 +(M^H)^2 \right)^{-1}$ in the two--point
function gives the wave--function renormalization $z_H$, and the
Ward identity \cite{10}
\bee
\rho_s^H={z_H \over \left( M^H \right)^4} H_s^2    \label{125}
\ee
gives then $\rho_s^H$. The wave--function renormalization $z_H$ has the form
of $z$ in eq.~(\ref{120}) replacing $D_{\infty}^{(0)}$ by
$D_{\infty}^{(M_0)}$. Eq.~(\ref{125}) leads to the result
\bee
\rho_s^H = \rho_s \left[ 1+2{N-2\over 8\pi\rho_s} M\right] \,. \label{126}
\ee
Expressing eq.~(\ref{123}) in terms of the renormalized parameters we get
finally
\bee
\left( M^H \right)^2 =M^2 \left[ 1- {N-3\over 8\pi}\cdot{M\over\rho_s}
\right] \,. \label{127}
\ee
After restoring dimensions, eqs.~(\ref{126}) and (\ref{127}) give the
result quoted in eq.~(\ref{9}).

The procedure above shows that the power divergences can be consistently
absorbed in the bare parameters, the physical predictions are free of
divergences as $a\to 0$, and they satisfy the Ward identities.

\bigskip
\section{\bf The effect of a weak staggered field}

Consider an euclidean box $L_s^{d-1}\times L_t$ with "cubic" geometry,
i.e. $L_t/L_s=$O$(1)$. If the staggered field is zero, the total
(staggered) magnetization moves freely in the O$(N)$ group space.
Switching on an external field ${\bf H}_s=(H_s,0,\ldots,0)$, the distribution
of the total staggered magnetization will be concentrated around the 0--th
direction. For weak fields
\bee
0 \le H_s = {\rm O} \left( {1 \over \Ms L_s^2 L_t}\right)
\label{7.1}
\ee
this distribution remains broad and there is a need to treat the
corresponding motion non--perturbatively.
The external field dependence
of the free energy, staggered magnetization and susceptibilities with
O$(N)$ symmetry have been worked out in ref.~\cite{10} up to next--to--leading
order. After restoring the spin wave velocity dependence we enumerated
the results in eqs.~(\ref{21}-\ref{26}).

Increasing the box size $L_s$ at fixed temperature, the staggered
susceptibility has the form
\bee
\chi_s(T,L_s) ={1\over N} \Ms^2 {L_s^2\over T}
\left\{ 1 +(N-1){T\over 2\pi\rho_s} \left( -\ln (L_sT) +{1\over 2} 2.852929
\right) + \ldots \right\}\,. \label{7.1a}
\ee
This behaviour remains valid until $L_s \ll \xi(T)$, where
$\xi(T) \propto \exp (b/T)$, with $b=2\pi\rho_s/(N-2)$
is the nonperturbatively generated
correlation length \cite{11,17}.
For infinite volume ($L_s \gg \xi(T)$) the spatial
size is expected to be replaced by the correlation length.

These results were derived under the condition $T=$O$(1/L_s)$ and, in general,
the $T\to 0$, $L_t=$fixed limit does not lead to the correct zero temperature
behaviour.
The physical reason was discussed in detail in Section 6.5. As we argued
there, the result of eq.~(\ref{26})
in this limit should agree with that given by the
high temperature limit of an O($N$) rotator. We shall discuss this problem
in the next section.

\section{\bf Low temperature limit of the finite volume staggered
susceptibility}

As we discussed in Section 6, if the temperature is very low ($T\ll 1/L_s$),
the excitations of an O($N$) rotator dominate the partition function.
Let us check first, how the rotator problem in eq.~(\ref{92C}) is
modified in the presence of a small staggered field
${\bf H}_s=(H_s,0,\ldots,0)$. In the effective action a new term enters
\bee
-\Ms^0 {\bf H}_s \cdot \sum_x {\bf S}_x \,,
\label{8.1}
\ee
where $\Ms^0$ is the bare staggered magnetization.
Using eqs.~(\ref{54}--\ref{61}) this term can be written as
\bee
-\Ms^0 V_s \sum_t {\bf H}_s {\bf e}(t) {1\over V_s}
\sum_{\vx} \left( 1- \vpi^2(x) \right)^{1/2} \,.
\label{8.2}
\ee
Considering the first correction we can replace $\vpi^2(x)$ by
$(1/\rho_s^0)(N-1)D^*(0)$, and eq.~(\ref{8.2}) leads to a new term in the
exponent of eq.~(\ref{92C}):
\bee
{1\over \kappa}\int d\e (t) \exp \left\{ -{1\over 2} \Theta_{eff}
\left( \e(t)-\e(t-1) \right)^2 +{\bf h}_s \e(t) \right\} \,,
\label{8.3}
\ee
where
\bee
{\bf h}_s=(h_s,0,\ldots,0)\,,~~~h_s=H_s \Ms^{eff} V_s\,,~~~
\Ms^{eff}=\Ms^0 \left(1- {N-1\over 2 \rho_s^0}D^*(0) \right) \,.
\label{8.4}
\ee
The corresponding Hamilton operator reads
\bee
\hat{H}={1\over 2 \Theta_{eff}} \hat{{\bf L}}^2 -{\bf h}_s\hat{{\bf X}}
=\hat{H}_0-{\bf h}_s\hat{{\bf X}},~~~\hat{{\bf X}}^2=1 \,,
\label{8.5}
\ee
where we suppressed the $h_s$--independent constant $E_0(L_s)$ of
eq.~(\ref{E1}).
Notice that this effective rotator picture is valid only in first order,
i.e. including O($1/\rho_s L_s$) corrections. This is the order considered
in this section.
The staggered susceptibility is given by
\bee
\chi_s(T,L_s)= \left. {T\over V_s}
{\partial^2 \over \partial H_s^2} \ln Z \right|_{H_s=0} =
\left. \left(\Ms^{eff} \right)^2 V_s T
{\partial^2 \over \partial h_s^2} \ln Z \right|_{h_s=0} \,,
\label{8.6}
\ee
where
\bee
Z={\rm Tr}\left( \rme^{-\hat{H}/T} \right) \,.
\label{8.7}
\ee
In taking the derivatives in eq.~(\ref{8.6}) we have to be careful since
$\hat{X}$ does not commute with $\hat{H}_0$. One obtains
\bee
\chi_s=\left(\Ms^{eff} \right)^2 V_s {1\over Z_0}
{\rm Tr}\int_0^{1/T} d\lambda \rme^{-\lambda \hat{H}_0} \hat{X}_0
\rme^{-(1/T-\lambda) \hat{H}_0} \hat{X}_0 \,,
\label{8.8}
\ee
where $Z_0$ is the partition function with
$\hat{H}_0$, and $\hat{X}_0$ is the $0$--th component of $\hat{{\bf X}}$.

The excitations of $\hat{H}_0$ are described by the quantum number
$j=0,1,\ldots$
\bee
E_j={1\over 2\Theta_{eff}} j(j+N-2) \,,
\label{8.9}
\ee
while the corresponding multiplicity is given by
\bee
g_j={(j+N-3)!\over j!(N-2)!}(2j+N-2) \,.
\label{8.10}
\ee
We shall denote the set of quantum numbers which characterize the
subspace with a given $j$ by $m$. Inserting a complete set of states in
eq.~(\ref{8.8}) we get
\bee
\chi_s=\left(\Ms^{eff} \right)^2 V_s {2\over Z_0}
\sum_{j=0}^{\infty}{\rme^{-E_j/T} -\rme^{-E_{j+1}/T}\over E_{j+1}-E_j} a_j \,,
\label{8.11}
\ee
where
\bee
a_j=\sum_{m} \left| \langle j+1,m | \hat{X}_0 | j,m \rangle \right|^2 \,.
\label{8.12}
\ee
In deriving eq.~(\ref{8.11}) we used that $\hat{X}_0$ connects the state
$|j,m\rangle$ with $|j\pm 1,m\rangle$ only.
Using the relation
\bee
\sum_m \sum_{j',m'} \left| \langle j',m' | \hat{X}_0 |j,m \rangle \right|^2=
\sum_m \langle j,m |\hat{X}_0^2 | j,m \rangle ={1\over N} g_j
\label{8.13}
\ee
and observing that the left hand side of eq.~(\ref{8.13}) is just
$a_j+a_{j+1}$, we obtain
\bee
a_j={1\over N}{(j+N-2)!\over j!(N-2)!} \,.
\label{8.14}
\ee
Writing eq.~(\ref{8.14}) into eq.~(\ref{8.11}), after some algebra we get
\beea
\chi_s&=&{4\over N(N-1)}\left(\Ms^{eff} \right)^2 \Theta_{eff} V_s {1\over Z_0}
\sum_{j=0}^{\infty} g_j \rme^{-E_j/T}
{(N-1)(N-3)\over 8\Theta_{eff}E_j +(N-1)(N-3)} \nonumber \\
 &=& {4\over N(N-1)}
\left(\Ms^{eff} \right)^2 \Theta_{eff} V_s {1\over Z_0} U(t) \,,
\label{8.15}
\eea
where $t=1/(2\Theta_{eff}T)$ and $U(t)$ satisfies the differential equation
\bee
{\partial U \over \partial t}=
{(N-1)(N-3)\over 4} \left[ U(t) - Z_0(t) \right] \,.
\label{8.16}
\ee
For $N=3$ eq.~(\ref{8.15}) simplifies since in this case $U(t)\equiv 1$.

Let us consider the effective parameters $\Ms^{eff}$ and $\Theta_{eff}$
of the O($N$) rotator. Using the definitions,
eqs.~(\ref{89},\ref{92A},\ref{8.4}) and replacing the bare spin stiffness
and magnetization by their renormalized values, eqs.~(\ref{116},\ref{122})
we have
\beea
\Ms^{eff}&=&\Ms \left[ 1+{N-1\over 2\rho_s}\left( D(0)-D^*(0)\right)\right]\,,
  \label{8.17A} \\
\Theta_{eff}&=&\rho_s V_s \left[ 1+{N-2\over \rho_s}
\left(D(0)-D^*(0)\right)+
{a\over\rho_s}\left({a^{1-d}\over 2d}+\Delta_0 D^*(0)\right)\right] \,,
\nonumber
\eea
where $D(0)$ is the lattice propagator at infinite volume and zero temperature.
In eq.~(\ref{8.17A}) we restored the lattice unit dependence to see clearly
how the ultraviolet divergences cancel. As eqs.~(A7,A14) in Appendix A
show, $D(0)$, $D^*(0)$ and $\Delta_0 D^*(0)$ are ultraviolet divergent.
On the other hand, the combinations
\beea
D(0)-D^*(0)&=& {1\over 4\pi L_s^{d-2}}
\left( - \alpha_{1/2}^{(d-1)}(1) +2 {d-1\over d-2} \right)
\nonumber \\
{a^{1-d}\over 2d}+\Delta_0 D^*(0) &=& {1\over 2 V_s} \label{8.20}
\eea
entering the effective parameters are finite.

Eqs.~(\ref{8.15}--\ref{8.16}) allows us to calculate $\chi_s$ both
for $t\ll 1$ and $t\gg 1$ analytically.
(Numerically, of course, it is not difficult to perform the calculation
for arbitrary $t$.)
At very low temperatures, the $j=0$ term contributes only in
eq.~(\ref{8.15}) and in $Z_0(t)$, giving
\bee
\chi_s(T,L_s)={ 4\over N(N-1)} \left(\Ms^{eff} \right)^2 \Theta_{eff} V_s
\left[ 1+{\rm O}\left( \rme^{-1/T\Theta_{eff}}\right) \right] \,,
{}~~~TV_s\rho_s \ll 1 \,.
\label{8.17}
\ee
or
\beea
& &\chi_s(T,L_s)={ 4\over N(N-1)} \Ms^2\rho_s V_s^2 \left[1+
{2N-3\over 4\pi\rho_s L_s^{d-2}}
\left( - \alpha_{1/2}^{(d-1)}(1) +2 {d-1\over d-2} \right)
\right] \,, \nonumber \\
& & T V_s\rho_s \ll 1\,. \label{8.21}
\eea
For the relevant case, $d=3$ we get
\bee
\chi_s(T,L_s)={ 4\over N(N-1)} \Ms^2\rho_s L_s^4 \left[1+
{2N-3\over 4\pi\rho_s L_s}3.900265 \right]\,,
 ~~~ T L_s^2\rho_s \ll 1\,,~~~d=3 \,.
 \label{8.22}
\ee
At large temperatures $T V_s\rho_s \gg 1$ (but $TL_s \ll 1$) we can use
Seeley's expansion eq.~(\ref{E3}) and integrate the differential equation
in eq.~(\ref{8.16}). We obtain
\beea
U(t) &=&  A \, {N-1\over 2} \, t^{-(N-3)/2}
\left[ 1+{(N-1)(N-3)\over 6}\, t + \right. \nonumber \\
&  & ~~~ \left.  {(N-1)(N-3)(5N^2-22N+18)\over 360}\,t^2
+ \ldots \right] \,,  \label{8.22A} \\
{U(t)\over Z_0(t)} &=& {2\over N-1} \, t \left[ 1-{N-1\over 6}\, t
+{(N-1)(N+1)\over 180}\, t^2 +\ldots \right] \,. \nonumber
\eea
Up to leading order corrections this leads to
\bee
\chi_s(T,L_s)={ 1\over N} \left( \Ms^{eff} \right)^2 {V_s \over T}
\left[ 1-{N-1\over 12 T \Theta_{eff}}\right] \,.
\label{8.23}
\ee
For $d=3$ we get
\bee
\chi_s(T,L_s)={ 1\over N} \Ms^2 {L_s^2\over T}
\left[ 1-{N-1\over \rho_s L_s }
\left(-{1\over 12}{1\over L_s T} +{1\over 4\pi} 3.900265\right)+
\ldots\right]\,. \label{8.24}
\ee

According to eq.~(\ref{8.22}), in the limit $T\to 0$, $V_s=$fixed,
the staggered susceptibility becomes independent of $T$
and is proportional to the square of the spatial volume. As the temperature
is increased to $T \gg \Theta =\rho_s V_s$, but $T \ll 1/L_s$, the rotator
becomes highly excited, while the fast Goldstone modes (whose typical
energy is $\sim 1/L_s$) remain frozen.
In this region the susceptibility
in eq.~(\ref{8.24}) should match the {\em low} temperature limit
of eq.~(\ref{26}).
Indeed, using the asymptotic expressions for $L_t \gg L_s$
given in Appendix B, one finds that, in the order considered in this section,
the two expressions coincide up to O($\exp (-\pi /L_s T)$) corrections,
coming from the fast modes  which are included in the cubic geometry.
It is interesting to note that the susceptibility given by the effective
rotator, eq.~(\ref{8.24}) reproduces the leading correction in eq.~(\ref{26})
surprisingly well even at temperatures as high as $T\sim 1/L_s$.
(See the discussion in Appendix B.)

\section{\bf Uniform susceptibility }

The uniform magnetic field $h_u$ in eq.~(\ref{2}) is coupled to the
generator of the O($3$) rotation around the direction of ${\bf h}_u$.
For general $N$, $h_u$ will be coupled to an O($N$) generator $Q$,
say $Q_{12}$ generating a rotation in the 1--2 plane.
We have $(Q{\bf S})_1=i S_2$, $(Q{\bf S})_2=-i S_1$,
the other components of $Q{\bf S}$
being zero. In the effective Lagrangean formulation, eq.~(\ref{30}),
the uniform field $h_u$ enters as the time component of a constant
imaginary gauge potential:
\bee
\partial_0 \to D_0 = \partial_0 - h_u Q \,. \label{9.1}
\ee
The effective Lagrangean in eq.~(\ref{30}) receives an extra contribution
\bee
\delta {\cal L}_u = -h_u \rho_s \left( \partial_0 {\bf S}(x) Q{\bf S}(x)
\right)
+{1\over 2} h_u^2 \rho_s^2 \left( Q{\bf S}(x) \right)^2 \,. \label{9.2}
\ee
For small $h_u$ it is again important to treat properly the zero modes.
Let us consider first the "cubic" geometry.
Introducing the global rotation $\Omega$: ${\bf S}(x)=\Omega \R(x)$  one has
to average $\exp(-\int \delta {\cal L}_u )$ over the O($N$) rotations $\Omega$.
This gives a contribution to the free energy
\beea
& &\delta f_u = -{1\over N} h_u^2 \rho_s \label{9.3} \\
& & -{2\over N(N-1)} h_u^2 \rho_s^2
 \left\langle {1\over V} \int_x \int_y
\left( \partial_0 \R(x) \R(y) \right)\left( \partial_0 \R(y) \R(x) \right)
\right\rangle_{h_u=0} +{\rm O}\left( h_u^3 \right) \,. \nonumber
\eea
Here we have used the following relations for averaging over the group
space:
\beea
& & \langle a_1' b_1' \rangle_{\Omega} = {1\over N}
({\bf a}{\bf b})\,,
\label{9.4} \\
& & \langle (a_1' b_2' -
a_2' b_1')(c_1' d_2' -
c_2' d_1') \rangle_{\Omega}
={2\over N(N-1)} \left[ ({\bf a}{\bf c})
({\bf b}{\bf d})-
({\bf a}{\bf d})({\bf b}{\bf c})
\right] \,, \nonumber
\eea
where ${{\bf a}'}=\Omega {\bf a}$, ${{\bf b}'}=\Omega {\bf b}$,
\ldots are rotated O($N$) vectors.
The expectation value in eq.~(\ref{9.3}) can be evaluated using
standard steps of chiral perturbation theory.
One obtains for the uniform
susceptibility at $h_u=0$ the result quoted in eq.~(\ref{27d}).

In eq.~(\ref{27d}) a new shape coefficient appears defined as
\bee
{1\over L^2} \psi(l)= -\int_V d^3x \partial_0^2 \tilde{G}(x) \tilde{G}^2(x)\,.
\label{9.5a}
\ee
Here $\tilde{G}(x)=\bar{G}(x)+\beta_1(l)/L
=1/(4\pi |x|)+{\rm O}(x^2)$, where
$\bar{G}(x)$ is defined in eq.~(\ref{46}).
For $l=1$ $\psi(l)$ is related to $\beta_n(1)$ (eq.~(\ref{27f})),
while for small $l$ its asymptotic form is given by
\bee
\psi(l) = -{1\over 4\pi^2 l^4}  +{\rm O} \left({\rm e}^{-2\pi/l^3}
\right),~~~ l \ll 1\,. \label{9.5b}
\ee
For generic values of $l$ it should be evaluated numerically
using methods discussed in detail in refs.~\cite{Gerber,Hansen}.
In Appendix~B we give approximate analytic expressions for $\psi(l)$.

A small staggered field $H_s$ will fix the direction
of the staggered magnetization provided $H_s \Ms L_s^2 \gg T$.
Relative to this direction we can define the transversal and longitudinal
uniform susceptibilities, $\chi_u^{\bot}(T,L_s)$ and $\chi_u^{\|}(T,L_s)$,
respectively.\footnote{For the general O($N$) case "parallel" is defined
by the O($N$) generators under which the staggered magnetization is invariant.}
The susceptibility $\chi_u(T,L_s)$ given in eq.~(\ref{27d}) corresponds to
$H_s=0$, and is given by the "angular" average over the O($N$) group space
\bee
\chi_u(T,L_s)={2\over N(N-1)}\left[ (N-1) \chi_u^{\bot}(T,L_s)+
{(N-1)(N-2)\over 2}\chi_u^{\|}(T,L_s) \right]\,. \label{9.5c}
\ee
In the infinite volume limit $\chi_u^{\bot}(T)$ and $\chi_u^{\|}(T)$
are infrared divergent in chiral perturbation theory. On the other hand,
the O($N$) invariant average $\chi_u(T)$ remains finite and it is given
by eq.~(\ref{27g}).

Small temperatures $T\ll 1/L_s$ should be considered again separately.
The Hamiltonian for the effective O($N$) rotator in this case is
\bee
\hat{H}={1\over 2 \Theta_{eff}} \hat{\bf L}^2
-h_u \hat{\bf L}_z \,, \label{9.7}
\ee
where we denoted for simplicity by $\hat{\bf L}_z$ the O($N$) generator
coupled to $h_u$.
With the notations of Section 8 we obtain
\bee
\chi_u (T,L_s)= -{2\over N(N-1)} {1\over V_s T}{1\over Z_0(t)}
{\partial \over \partial t}Z_0(t) \,. \label{9.8}
\ee
For $TV_s\rho_s \gg 1$ (but $T \ll 1/L_s$) Seeley's expansion gives
\bee
\chi_u (T,L_s)={2\over N}\rho_s \left\{ 1+{N-2\over \rho_s L_s^{d-2}}
\left[ -{1\over 6L_sT}+{1\over 4\pi}\left(-\alpha_{1/2}^{(d-1)}(l)
+2{d-1\over d-2}\right) \right] \right\} \,, \label{9.9}
\ee
in agreement with the low temperature behaviour obtained from
eq.~(\ref{27d}).

For very low temperatures, $TV_s\rho_s \ll 1$, only the lowest lying
rotator states contribute, leading to an exponentially small susceptibility
given by eq.~(\ref{27c}).

{\bf Acknowledgements} The authors are indebted to Wolfgang Bietenholz, Heping
Ying, Heiri Leutwyler and Uwe Wiese for discussions. A correspondence
with Ted Barnes and Dan Arovas helped to clarify some problems, which we
gratefully acknowledge.
Parts of this work have been completed in the Max--Planck--Institute
for Physics in Munich. The authors thank for the kind hospitality.

\appendix
\section{\bf Momentum sums and integrals on the lattice}

It is useful to introduce the notations
\beea
& & \hat{k}_{\mu}=\rme^{ik_{\mu} }-1\,,
{}~~~\hat{k}_{\mu}^*=\rme^{-ik_{\mu} }-1\,,~~~
d(k)=4\sum_{\mu=1}^4 \sin^2{k_{\mu}\over 2} \,, \nonumber \\
& & \int_k = \int_{-\pi}^{\pi} \ldots \int_{-\pi}^{\pi}
{d^d k\over (2\pi)^d }\,,~~~
{\sum_k}^* =\sum_{k\atop k^0 \neq 0} \,,~~~
{{\sum}_k}' =\sum_{k \atop k \neq (0,\ldots,0)} \,, \label{A1} \\
& & D^*(x)={1\over V} {\sum_k}^* {\rme^{ikx}\over d(k)}
\eea
With the help of the identities
\bee
-\left(\hat{k}_{\mu}+\hat{k}_{\mu}^*\right)=\hat{k}_{\mu}\hat{k}_{\mu}^*=
4 \sin^2{k_{\mu}\over 2}\,,~~~({\rm no~sum~over~} \mu) \,, \label{A2}
\ee
one obtains
\beea
& &\left. \Delta_{\mu}^x\Delta_{\mu}^y D^*(x-y) \right|_{x=y}
= {1\over V} {\sum_k}^* {\hat{k}_{\mu}\hat{k}_{\mu}^*\over d(k)}
\label{A3} \\
& &~~~ = -{1\over V} {\sum_k}^* {\hat{k}_{\mu}+\hat{k}_{\mu}^*\over d(k)}
= -2 \Delta_{\mu} D^*(0)\,.~~~({\rm no~sum~over~} \mu )
\eea
In the following we shall discuss the volume dependence of $D^*(0)$
and $\Delta_{\mu}D^*(0)$ in a euclidean box $L_s^{d-1}\times L_t$
in the $L_t \to\infty$ limit.

The sum in $D^*(0)$ can be written as
\bee
D^*(0)={1\over V}{\sum_k}^* {1\over d(k)} =
\lim_{M^2\to 0} \left\{ {1\over V} \sum_k {1\over d(k)+M^2}
-{1\over V} \sum_{k^0}{1\over d(k^0,\vec{k}=0)+M^2}\right\}\,. \label{A4}
\ee
We divide the sums in eq.~(\ref{A4}) into the infinite volume
contribution and the rest, which is volume dependent and remains finite
as the lattice unit $a\to 0$:
\bee
D^*(0)=\lim_{M^2\to 0} \left\{ \int_k {1\over d(k)+M^2} +g_1^{(d)}
-{1\over V_s} \left[ \int_{k_0}{1\over 4\sin^2{k_0\over 2}+M^2}
+g_1^{(d=1)} \right] \right\} \,. \label{A5}
\ee
Since we are not interested in cut--off effects which go to zero as the
cut--off goes to infinity, we can evaluate $g_1^{(d)}$ in the $a\to 0$
continuum limit. This problem is discussed in detail in the Appendix B
of ref.~\cite{10} and we quote the result only:
\beea
\left. g_1^{(d)}\right|_{{M\to 0 \atop L_t \to \infty}}
      &=& {1\over 4\pi L_s^{d-2}}
\left[ \alpha_{1/2}^{(d-1)}(1)+{4\pi\over 2ML_s} -2{d-1 \over d-2} \right]\,,
\nonumber \\
\left. g_1^{(1)}\right|_{L_t\to\infty} &=& 0 \,. \label{A6}
\eea
In the $M\to 0$ limit the first integral in eq.~(\ref{A5}) is finite while
the second one gives $1/2M$. We obtain
\bee
\left. D^*(0)\right|_{L_t\to\infty} =\int_k {1\over d(k)}
+{1\over 4\pi L_s^{d-2}}
\left[ \alpha_{1/2}^{(d-1)}(1)-2{d-1\over d-2} \right] \,. \label{A7}
\ee
The first term is a volume independent but cut--off dependent constant,
while the number $\alpha_{1/2}^{(d-1)}(1)$ can be found in Table~1 in
ref.~\cite{10}.

The steps are similar for $\Delta_{\mu}D^*(0)$:
\beea
& & \Delta_{\mu}D^*(0)={1\over V} {\sum_k}^*{\hat{k}_{\mu}\over d(k)}
  \label{A8} \\
& & ~~~ = \lim_{M^2\to 0} \left\{ {1\over V}
\sum_k {\hat{k}_{\mu}\over d(k)+M^2}
-\delta_{\mu 0}{1\over V_s}{1\over L_t}
\sum_{k^0}{1\over d(k^0,\vec{k}=0)+M^2}\right\}\,. \nonumber
\eea
The second term in eq.~(\ref{A8}) gives $+\delta_{\mu 0}/2V_s$.
In the first term we separate the infinite volume contribution:
\bee
\lim_{M^2\to 0} {1\over V}\sum_k {\hat{k}_{\mu}\over d(k)+M^2} =
{1\over 2}\int_k {\hat{k}_{\mu}+\hat{k}_{\mu}^*\over d(k)} +Q_{\mu} \,,
\label{A9}
\ee
where
\bee
Q_{\mu}= \lim_{M^2\to 0}  {1\over 2}
\left[ {1\over V}{\hat{k}_{\mu}+\hat{k}_{\mu}^*\over d(k)+M^2}
-\int_k {\hat{k}_{\mu}+\hat{k}_{\mu}^*\over d(k)+M^2} \right] \,.
\label{A10}
\ee
It is easy to show that
\bee
Q_i=-{1\over d-1} Q_0,~~~~i=1,\ldots,d-1\,. \label{A11}
\ee
Indeed, for $L_t \to\infty$ we can write
\beea
Q_0 &=& {1\over 2} \left( {1\over V_s}\sum_{\vec{k}}-\int_{\vec{k}}\right)
\int_{-\pi}^{\pi}{dk_0\over 2\pi}{-4\sin^2{k_0\over 2}\over d(k)}
\label{A12} \\
  &=& {1\over 2} \left( {1\over V_s}\sum_{\vec{k}}-\int_{\vec{k}}\right)
\int_{-\pi}^{\pi}{dk_0\over 2\pi}{\sum_i 4\sin^2{k_i\over 2}\over d(k)}
=-\sum_i Q_i \,. \nonumber
\eea
Since we have a box $L_s^{d-1}\times L_t$, $Q_i$ is independent of $i$
and eq.~(\ref{A11}) follows. The first term in eq.~(\ref{A9}) gives
\bee
{1\over 2}\int_k {\hat{k}_{\mu}+\hat{k}_{\mu}^*\over d(k)}=
-{1\over 2d}{1\over a^{d-1}} \,, \label{A13}
\ee
where on the right hand side we restored the cut--off dependence
explicitly. We have, therefore
\beea
& & \Delta_0 D^*(0)= -{1\over 2d}{1\over a^{d-1}}+{1\over 2V_s} +Q_0 \,,
\label{A14} \\
& & \Delta_i D^*(0)= -{1\over 2d}{1\over a^{d-1}}-{1\over d-1} Q_0\,.
\nonumber
\eea
In the expression for the ground state energy eq.~(\ref{102}) the combination
\beea
\Delta_{\mu}D^*(0)\Delta_{\mu}D^*(0)-{1\over V_s}\Delta_0 D^*(0)=
\left(-{1\over 2d}{1\over a^{d-1}}+{1\over 2V_s} +Q_0\right)^2 \label{A15} \\
+(d-1)\left( -{1\over 2d}{1\over a^{d-1}}-{1\over d-1} Q_0\right)^2
-{1\over V_s}\left(-{1\over 2d}{1\over a^{d-1}}+{1\over 2V_s} +Q_0\right)
\nonumber
\eea
enters. Eq.~(\ref{A15}) leads to the result
\bee
\Delta_{\mu}D^*(0)\Delta_{\mu}D^*(0)-{1\over V_s}\Delta_0 D^*(0)=
{1\over 4d}{1\over a^{2d-2}}+{d\over d-1}Q_0^2-{1\over 4V_s^2} \,.\label{A16}
\ee
As expected, there is no mixing between divergent cut--off powers and volume
dependence in eq.~(\ref{A16}). Since $Q_0$ is finite in the $a\to 0$ limit,
we can write
\beea
Q_0&=&  \lim_{M^2\to 0} \left( -{1\over 2} \right)
\left[ {1\over V_s} \sum_{\vec{k}}
-\int_{\vec{k}}\right] \int_{-\infty}^{\infty}{dk_0\over 2\pi}
{k_0^2\over k_0^2+\vec{k}^2} \label{A17} \\
 &=& {1\over 2}\left[ {1\over V_s} \sum_{\vec{k}}\left(\vec{k}^2\right)^{1/2}
-\int_{\vec{k}}\left(\vec{k}^2\right)^{1/2} \right] =
\left. -{1\over 4\sqrt{\pi}} g_{-1/2}^{(d-1)} \right|_{M^2\to 0} \,, \nonumber
\eea
where, in the last step, we used the notation of ref.~\cite{10}.
Since $g_{-1/2}^{(d-1)} \sim L_s^{-d}$, after restoring the dimensions
we get
\bee
Q_0={\rm O}\left({a\over L_s^d}\right) \,, \label{A18}
\ee
and $Q_0^2$ can be neglected in eq.~(\ref{A16}) in our order.
We can write therefore
\bee
\left. \Delta_{\mu}D^*(0)\Delta_{\mu}D^*(0)-{1\over V_s}\Delta_0 D^*(0)
\right|_{{L_t\to \infty \atop a\to 0}} =
{1\over 4d}{1\over a^{2d-2}}-{1\over 4V_s^2} \,. \label{A19}
\ee
The divergent first term contributes to the non--universal constant
of the ground state energy density only.

\section{\bf Asymptotic expressions}

For the "slab geometry", $l \ll 1$ (or $L_t \ll L_s$) and for
the "cylinder geometry", $l \gg 1$ ($L_t \gg L_s$) one can derive
simple asymptotic expressions for the functions
$\alpha_r(l)\equiv \alpha_r^{(3)}(l)$, which are related by eq.~(\ref{14a})
to the functions $\beta_n(l)$, $n=0,1,2$
appearing in the finite size corrections.
One can show that for $l\leq 1$
\beea
\alpha_r^{(3)}(l) &=& l^{4r-6}\left[ \alpha_{-r+3/2}^{(1)}(1)
+{1\over (r-1)(2r-3)}\right]
+l^{-2r}\left[ \alpha_{r}^{(2)}(1)+{1\over r(r-1)}\right]
\nonumber \\
& & -{3\over r(2r-3)}+{\rm O}\left( \rme^{-\pi/ l^2}\right) \,,
{}~~~l\leq 1\,. \label{B01}
\eea
For $l \geq 1$ one has the approximation:
\beea
\alpha_r^{(3)}(l) &=& l^{4r}\left[ \alpha_r^{(1)}(1)+{1\over r(2r-1)}\right]
+l^{3-2r}\left[ \alpha_{-r+3/2}^{(2)}(1)+{4\over(2r-1)(2r-3)}\right]
\nonumber \\
& & -{3\over r(2r-3)}+{\rm O}\left( \rme^{-\pi l^2}\right) \,,
{}~~~l\geq 1\,. \label{B0}
\eea

Using these expressions and the equality
$\alpha_p^{(d)}(1) = \alpha_{d/2-p}^{(d)}(1)$ one can relate the shape
coefficients to those corresponding to a symmetric box in $d$ dimensions,
listed in Table~1 of ref.~\cite{10}.

In particular we have for $l \leq 1$
\beea
\beta_0 (l) &=& l^{-6}\cdot 0.382626 +2\ln l +1.054689 +
\delta \beta_0^- (l)\,, \nonumber \\
 {1\over l}\beta_1(l) &=& {1\over 4\pi l^3}
\left[ 2.852929 + 6 \ln l \right]
+ {1\over l}\delta\beta_1^-(l) \,, \nonumber \\
 {1\over l^2}\beta_2(l) &=& {1\over (4\pi)^2}
\left[ { 0.610644\over l^6} + 1.047198 \right]
+ {1\over l^2}\delta\beta_2^-(l)  \,.
\label{B1}
\eea
For $l\geq 1$ we get
\beea
\beta_0 (l) &=& l^3\cdot 1.437745 -4\ln l +
 \delta \beta_0^+ (l) \,, \nonumber \\
 {1\over l}\beta_1(l) &=& -{1\over 12}l^3 +{3.900265\over 4\pi}+
 {1\over l}\delta\beta_1^+(l)
\,, \nonumber \\
 {1\over l^2}\beta_2(l) &=& {1\over 720}l^6 +{1.437745\over (4\pi)^2 l^3}
+{1\over l^2}\delta\beta_2^+(l) \,.
\label{B2}
\eea
The correction terms $\delta\beta_n^{\mp}(l)$
are exponentially small for $l\ll 1$ and $l\gg 1$,
respectively. They can be parametrized in a simple form
\beea
& & \delta\beta_n^- (l)=a_n^- l^{b_n^-} {\rm e}^{-2\pi/l^3}\,,
{}~~~ l\le 1\,,\nonumber \\
& & \delta\beta_n^+ (l)=a_n^+ l^{b_n^+} {\rm e}^{-2\pi l^3}\,,
{}~~~l\ge 1 \,. \label{B3}
\eea
The exponential factors are expected on physical grounds, while the
parameters $a_n^{\pm}$, $b_n^{\pm}$ can be determined from
the known values of $\beta_n(1)$ and the fact that $\beta_n'(1)=0$.
They are summarized in Table 1.
It is remarkable that this simple parametrization gives a
relative error smaller than $10^{-5}$ in the whole range of $l$ values.

\begin{table}
\centering
\begin{tabular}{crrrr} \hline
$n$ & $a_n^-$~~~~ & $b_n^-$~~~~ & $a_n^+$~~~~ & $b_n^+$ ~~~~\\
\hline
0 & $8.85264$ & $-0.959427$ & $8.62200$ & $-0.608985$ \\
1 & $-0.66610$ & $-0.321512$ & $-0.671988$  & $0.553185$ \\
2 & $0.058438$  & $1.35468$ & $0.061051$ &  $1.25040$ \\
\hline
\end{tabular}
\caption{Values of the parameters in eq.~(\protect\ref{B3}).}
\end{table}

For the shape coefficients $\psi(l)$ defined in eq.~(\ref{9.5a})
one can also give approximate analytic expressions
\beea
& & \psi(l)=-{1\over 4\pi^2 l^4}+2.57 \, {\rm e}^{-2\pi/l^3}\,,
{}~~~ l\le 1\,, \label{B4} \\
& & \psi(l)=-{1\over 120} l^8 +{3.900265\over 24\pi} l^5
-0.055 \, l^2 -4.78 \, {\rm e}^{-2\pi l^3}\,,
{}~~~l\ge 1 \,. \nonumber
\eea

        \end{document}